\documentstyle[12pt]{article}            
\begin{document} 
\begin{center} 
\Huge 
{\bf Infrared and Optical Spectroscopy of Type~Ia Supernovae in the  
Nebular Phase} \\  
\end{center} 
\vspace{1cm} 
 
\large 
E.J.C. Bowers$^1$, W.P.S. Meikle$^1$, T.R. Geballe $^2$, N.A. Walton$^3$,  
P.A.~Pinto$^4$, V.S. Dhillon$^5$, S.B. Howell$^6$, M.K. Harrop-Allin$^7$. \\ 
 
\vspace {0.5cm} 
 
\small 
$^1$Astrophysics Group, Blackett Laboratory, Imperial College of Science, 
Technology and Medicine, Prince Consort Road, London SW7 2BZ, UK \\  
$^2$Joint Astronomy Centre, 660 N. A'ohoku Place, University Park, Hilo, Hawaii 
96720, USA \\  
$^3$Royal Greenwich Observatory, Apartado de Correos 321, 
38780 Santa Cruz de La Palma, Tenerife, Islas Canarias, Spain \\ 
$^4$Steward Observatory, University of Arizona, Tucson, AZ 85721, USA \\  
$^5$Royal Greenwich Observatory, Madingley Road, Cambridge 
CB3 0EZ, UK \\  
$^6$Department of Physics and Astronomy, University of Wyoming, PO Box 
3905, University Station, Laramie, WY 82071, USA \\ 
$^7$Mullard Space Science Laboratory, University College London,  
Holmbury St. Mary, Dorking, Surrey, RH5 6NT, UK \\ 
\date{Accepted 1997. Received 1997; in original form 1997}  
 
\newpage 
\normalsize 
\begin{abstract} 
We present near-infrared (NIR) spectra for Type~Ia supernovae at epochs 
of 13 to 338~days after maximum blue light.  Some contemporary optical 
spectra are also shown. All the NIR spectra exhibit considerable 
structure throughout the J-, H- and K-bands.  In particular they exhibit 
a flux `deficit' in the J-band which persists as late as 175~days. This 
is responsible for the well-known red J-H colour.  To identify the
emission features and test the $^{56}$Ni hypothesis for the explosion
and subsequent light curve, we compare the NIR and optical
nebular-phase data with a simple non-LTE nebular spectral model. We
find that many of the spectral features are due to iron-group elements
and that the J-band deficit is due to a lack of emission lines from
species which dominate the rest of the IR/optical spectrum.
Nevertheless, some emission is unaccounted for, possibly due to
inaccuracies in the cobalt atomic data. For some supernovae, blueshifts of 
1000--3000~km/s
are seen in infrared and optical features at 3~months. We suggest this
is due to clumping in the ejecta. The evolution of the cobalt/iron
mass ratio indicates that $^{56}$Co-decay dominates the abundances of
these elements. The absolute masses of iron-group elements 
which we derive support the basic thermonuclear
explosion scenario for Type~Ia supernovae.  A core-collapse origin is
less consistent with our data. \\
 
\end{abstract}

\newpage 
\section {Introduction}  
 
It is widely believed that the luminosity of a Type~Ia supernova (SNIa) 
is due to the deposition of gamma-rays and positrons from the 
radioactive decay, $^{56}$Ni $\rightarrow$ $^{56}$Co 
$\rightarrow$ $^{56}$Fe, of $^{56}$Ni created in the explosion (Pankey 
1962; Colgate \& McKee 1969).  Given the central importance of this 
process it is surprising that only a couple of pieces of {\it direct} 
evidence in favour of a radioactively driven light curve have been 
presented. Axelrod (1980a,b) studied the [Co~III]~0.59~$\mu$m feature in 
the Type~Ia SN~1972E, and found that its decline was consistent with 
$^{56}$Co decay. A similar result was obtained by Kuchner {\it et al.} 
(1994) who used optical spectra of a sample of Type~Ia SNe to follow 
the Co/Fe mass ratio. Both of these studies made use of optical 
spectroscopy of Type~Ia SNe in their `nebular' or `late-time' phase. By 
this, we mean the era starting about 60~days post-maximum-light when 
the ejecta are optically thin to most of the infrared (IR) and optical 
lines.  (Throughout this paper, epochs are defined with respect to 
maximum light in the B-band, t$_{Bmax}$=0~days). Nevertheless, some of 
the stronger optical lines can remain optically-thick for a significant 
time after the nebular phase begins. Clumping can exacerbate this 
effect.  In addition, the sheer number of strong lines in the optical 
region results in severe line blending, causing identification 
uncertainty and inaccurate flux measurement of individual lines. In 
particular, the identification and measurement of the 
[Co~III]~0.59~$\mu$m line was vital for the work of Axelrod and Kuchner 
{\it et al.}.  A concern about this identification was the extent of 
contamination by Na~I~D. Unfortunately, there are no other strong lines 
of Na~I present in the optical spectra and so it was difficult 
to place an observational constraint on the strength of Na~I~D. 
Consequently, some doubt was expressed (Wheeler {\it et al.} 1993) 
as to the reliability of the 0.59~$\mu$m feature as a direct test of the 
radioactive decay model.  However, as Kuchner {\it et al.} point out, 
it can be argued on theoretical grounds that significant Na~I~D 
contamination is unlikely. 
 
Hoyle \& Fowler (1960) proposed that a Type Ia supernova results from 
the explosion of a white dwarf due to the thermonuclear fusion to 
$^{56}$Ni of $\sim$0.5--1~M$_\odot$ of the nuclei suspended in the 
electron-degenerate gas.  Different classes of models which invoke this 
mechanism are the Chandrasekhar-mass models (e.g. Iben \& Tutukov 1984, 
Nomoto, Thielemann \& Yokoi 1984; Khokhlov 1991; Woosley \& Weaver 
1994a) and sub-Chandrasekhar-mass models (Livne 1990; Livne \& Glasner 
1990; Woosley \& Weaver 1994b; Livne \& Arnett 1995; H\"{o}flich \& 
Khokhlov 1995). While all these models tend to eject 0.2--1.0~M$_\odot$ 
of $^{56}$Ni, they nevertheless all have problems (Woosley \& Weaver 
1994b). 

The thermonuclear explosion mechanism is challenged by S.~Colgate and
collaborators (Colgate {\it et al.} 1997). They favour instead a
core-collapse model with low mass ejection (\(<\)0.1~M$_{\odot}$). The
reason they advocate this hypothesis is due to apparent difficulties in
accounting for the decline rates of the optical light curves at late
times. As early as 100 days it is observed that the Type~Ia BVR fluxes
decline with a faster time-scale than the 77~day half-life of $^{56}$Co.
In both the thermonuclear and core-collapse scenarios, this can be
initially explained as being due to an increasing fraction of the
gamma-rays escaping as the ejecta expand. Eventually, however, the
luminosity is driven predominantly by the positrons which are more easily
absorbed than the gamma-rays (positron emission occurs in 4\% of the
$^{56}$Co decays). In the thermonuclear case, the large ejecta mass
($\sim$1~M$_{\odot}$) means that the positrons continue to be totally
trapped, even after they dominate the luminosity.  The fact that the
light curve nonetheless continues to decline faster than the $^{56}$Co
decay rate is explained as being due to an increasing fraction of the
energy emerging at wavelengths longward of 2.5~$\mu$m and so escaping
detection. In particular, strong fine-structure lines of [Fe~II] and
[Fe~III] in the 15--30~$\mu$m range are anticipated (Axelrod, 1980). This
crucial prediction has yet to be tested.  As mentioned above, an
alternative explanation is provided by the small ejecta mass
($\sim$0.1~M$_\odot$) of the core-collapse model which results in a
steadily increasing transparency to the positrons. 

Clearly, it is vital to establish the mass of $^{56}$Ni in the SN ejecta
in order to distinguish between the thermonuclear and core-collapse
explosion mechanisms. In particular, a conclusive test would be to
determine the mass of iron in the ejecta at late times. Such a
measurement also holds out the prospect of distinguishing between
specific thermonuclear models, although higher precision is probably
needed here. The measurement of the absolute mass of ejected iron is
difficult and has rarely been attempted. Significant uncertainties exist
in distance, atomic data, and ejecta conditions (ionization,
temperature, density). The first attempts were by Meyerott (1980) and
Axelrod (1980a,b) who inferred a substantial mass of ejected $^{56}$Ni
from the forbidden iron lines in the optical spectra of the Type~Ia
SN~1972E, although the aforementioned uncertainties meant that only a
weak constraint on the mass (0.3--1.0M$_{\odot}$) was achieved. More
recently, Spyromilio {\it et al.} (1992) used near-infrared (NIR) and 
optical spectra
of the {\it peculiar} Type~Ia SN~1991T to deduce the presence of about
0.5~M$_\odot$ of $^{56}$Fe in the ejecta but also with a large
uncertainty (0.4--1.2M$_\odot$).  Ruiz-Lapuente (1992) modelled the
optical spectra of the {\it peculiar} Type Ia SN~1986G and derived a
$^{56}$Ni mass of about 0.4 M$_{\odot}$, although this too was
subject to large uncertainties in the atomic data used.

We conclude that the widespread belief in the standard Type~Ia scenario
of a thermonuclear explosion and radioactively-driven light curve rests
on remarkably little direct evidence.  In order to place our
understanding of SNIas on a more secure observational basis we have begun
a programme of NIR spectroscopy of a number of typical Type~Ia supernovae
during the nebular phase.  Since the ejecta are optically thin,
observation during this era offers the best prospects of measuring
directly total relative and absolute masses. In the nebular phase we
expect the ejecta to be cooled mainly through forbidden-line emission
from singly- and doubly-ionized cobalt and iron atoms.  Observation in
the NIR region has a couple of advantages over the optical band. Optical
lines are generally of higher excitation and so are more
temperature-sensitive.  In addition the opacity of certain optical
transitions are higher than those in the IR, and may even be optically
thick, especially if the ejecta are clumped. The presence of forbidden
NIR lines of iron and cobalt in the Type~Ia nebular-phase spectra
together with the advent of sensitive IR spectrographs, and significant
improvements in the accuracy of atomic data (Hummer {\it et al.} 1993),
offers the possibility of a powerful direct test of both the explosion
mechanism and of the radioactive decay hypothesis.

To test for the presence of radioactive cobalt we use forbidden cobalt 
and iron lines and a simple spectral synthesis model to determine both 
the total cobalt/iron mass {\it ratio} and the absolute masses of 
cobalt and iron over a sequence of epochs. The determination of 
absolute mass is, in principle, subject to greater uncertainty owing to 
its sensitivity to distance and temperature errors. The ratio method 
was applied originally to the Type~II SN~1987A by Varani~{\it 
et~al.}~(1990). The technique exploits several interesting features of 
the cobalt-iron nebula which render it relatively insensitive to the 
details of the spectral synthesis code. Firstly, the similar atomic 
configurations and thus excitation energies of iron and cobalt, plus 
the fact that if the iron and cobalt do arise from $^{56}$Ni decay then 
they should be spatially co-extensive in the ejecta, means that 
estimates of the {\it ratios} of masses of Co$^+$/Fe$^+$ or 
Co$^{2+}$/Fe$^{2+}$ are much less sensitive to temperature or 
temperature gradients than are absolute mass determinations (Varani 
{\it et al.} 1990). Secondly, the first ionization potentials of iron 
and cobalt are virtually the same (7.9~eV for Fe and 7.87~eV for Co), 
and the second ionization potentials differ by only 5\% (16.19~eV for 
Fe and 17.08~eV for Co) (Fuhr, Martin and Wiese 1988). Consequently the 
individual mass ratios Co$^+$/Fe$^+$ or Co$^{2+}$/Fe$^{2+}$ derived 
from model spectral matches to the data should be equal to the total 
cobalt/iron mass ratio. Thirdly, such ratios are largely unaffected by 
uncertainties in distance or extinction.  This is particularly so for 
infrared lines where the extinction is lower. 
 
Prior to this work, little was known about the NIR behaviour of Type~Ia 
SNe in the nebular phase. NIR photometry in the 60--100 day era (Elias 
{\it et al.} 1981, 1985) revealed a strong flux-deficit in the J-band 
(J--H$\sim$+1.5).  Indeed, this deficit was present as early as 
15~days. The first NIR spectra of Type~Ia SNe, {\it viz.} SN~1986G  
(Frogel {\it et al.} 1987; Graham {\it et al.} 1987) (published in 
Meikle {\it et al.} 1997), SN~1991T (Spyromilio, Pinto \& Eastman 
1994) and SN~1995D (Meikle {\it et al.} 1997) showed that the J--H 
reddening in the 13--60~day era is due to the presence of a deep, wide 
flux deficit between 1.1 and 1.5~$\mu$m. This deficit, originally 
suggested to be a broad absorption (Kirshner {\it et al.} 1973; Elias 
{\it et al.} 1981; Graham {\it et al.} 1987), has more recently been 
attributed to a lack of emission lines in that region (Spyromilio, 
Pinto \& Eastman 1994; Meikle {\it et al.} 1997). 
 
In this paper we present near-IR spectra of seven Type~Ia supernovae, 
namely SNe~1986G, 1991T, 1992G, 1994ae, 1995D, 1995al and 1996X. The 
92G, 94ae, 95al and 96X spectra are presented for the first time. These 
observations span epochs 13~days to 338~days, thus providing the most 
extensive temporal coverage ever achieved for Type~Ia SNe NIR spectra. 
While this work focuses on those spectra taken during the nebular 
phase only, for completeness we have also included 
several spectra taken during the earlier photospheric phases of some of 
these supernovae. Type~Ia NIR spectra in the photospheric phase have 
already been presented and discussed by Frogel {\it et al.} (1987), 
Lynch {\it et al.} (1990, 1992) and Meikle {\it et al.} (1996, 1997). 
For the nebular phase spectra, line identification and interpretation 
is carried out using our non-LTE, nebular spectral synthesis code. We 
shall use this analysis to (a) identify the lines (b) investigate the 
origin of the J-band deficit, (c) determine the evolution of the 
cobalt/iron mass ratio, and (d) estimate the absolute mass of $^{56}$Ni 
ejected. 
 
\section {Observations}
Descriptions of the seven Type~Ia supernovae under consideration are
given in Table~1. Infrared spectra, taken at ten epochs spanning 13 to
338~days post-maximum blue light are shown in Figures~1 \& 2. The
observing log for the NIR spectra is in Table~2. The SN~1986G spectra
were acquired with CGS2 at UKIRT, while the day~60 spectrum of SN~1991T
was obtained with IRIS at the Anglo-Australian Telescope (AAT). All the
other infrared spectra were acquired using the cooled grating
spectrograph, CGS4, at the United Kingdom Infrared Telescope (UKIRT) with
the short focal length (150~mm) camera and either the 75 or 150~l/mm
grating. Sky subtraction was achieved by nodding between two positions
along the slit. Data reduction was carried out using the CGS4DR (Daly
1996) and Figaro (Shortridge 1995) packages. Except where stated
otherwise, the CGS4 wavelength calibration was with respect to krypton
and argon arc lamps. Where roughly contemporary optical spectra are
available, these are shown in Fig.~3.  The observing log for the optical
spectra is given in Table~3. 

Since our aim was to simultaneously model contemporary NIR {\it and}
optical spectra, close attention was given to both relative and
absolute fluxing. Only for SN~1995D was there a sufficiently high
signal-to-noise overlap in the optical and NIR spectra (0.85--1.04~$\mu$m) 
to allow the relative fluxing to be directly checked.  It was found
that the total flux in the overlap range as measured from the NIR and
optical spectra differed by only 1\%.  This suggests not only that the
relative fluxing was good but also, given that the NIR and optical data
were obtained at different times, at different sites and with different
instruments, it gives us confidence about the quality of our absolute
fluxing.

All spectra were absolute flux-calibrated using flux standards. In
addition, the fluxing was checked by comparing our observed spectral
fluxes in the standard pass-bands with those derived by extrapolation
from photometry, using Type~Ia light curves. Fluxing in the
optical band was checked by extrapolating early-epoch V-band photometry
to the epochs of our observations, using the standard V-band light
curve of Doggett \& Branch (1985).  These magnitudes were then compared
with our observed total spectral flux in a square passband spanning
0.5055--0.5945~$\mu$m.  For SN~1995D at 92~d and 1995al at 176~d the
broadband flux obtained by extrapolation exceeded the observed total
spectral flux by about 30\%.  For SN~1991T, the difference was less
than 5\%. 

Absolute fluxing in the NIR band was checked as follows. Using IRCAM2
at UKIRT, H- and K-band photometry of SN~1995al was obtained at 162~d.
No NIR light curves are available for this era and so we used the
I-band light curve of Schlegel (1995) to extrapolate to 94~d and 175~d.
The extrapolated fluxes were then compared with the total observed
spectral fluxes in square passbands spanning 1.45--1.85~$\mu$m and
1.9--2.5~$\mu$m. We found that the spectral fluxes tended to
exceed the extrapolated broadband fluxes by about 15~\%. For SN~1994ae
at 174~d we followed a similar procedure, extrapolating from AAT/IRIS
H-photometry obtained at 120~days (courtesy of J. Spyromilio). In this
case, the spectral flux was fainter than the extrapolated value by
$\sim$40\%, but the signal-to-noise is very low. For SN~1995D, no NIR
photometry was available.  We therefore scaled the observed NIR
spectral flux by an amount equal to the difference in the peak
V-magnitudes of SNe~1995D and 1995al.  This was then compared with the
extrapolated H-band flux as for SN~1995al. We found the observed
spectral flux was less than the extrapolated broadband value by 15\%.
Only for the NIR fluxing of SN~1991T at 338~d was a photometry check
not possible due to the non-availability of reliable infrared light
curves extending to such a late epoch.

Given that the square passbands we adopted only roughly approximate the
shape of the photometry passbands used, we conclude that the agreement
between the measured spectral fluxes and the extrapolated values is
encouraging.  It seems unlikely that the fluxing of either the optical
or NIR spectra was in error by more than $\sim$25\%.  Such an error is
small compared with other modelling uncertainties discussed later.

\subsection{SN~1986G} 
SN~1986G occurred in the dust lane of NGC~5128 (Cen~A). Its B--V colour 
(Hill 1986; Phillips and Geisler 1986), ultraviolet spectra (Wamsteker 
{\it et al.} 1986; Kirshner 1986) and Na~I~D and Ca~II H,K absorption 
lines observed at the redshift of the parent galaxy (Tonry and Strauss 
1986; Heathcote, Cowley and Hutchings 1986) all indicate that the 
supernova was substantially reddened. In addition infrared light curves 
showed that SN~1986G was intrinsically atypical (Frogel {\it et al.} 
1987).  Spectra spanning $\sim$1--2.5~$\mu$m for the period 13 to 
30~days were acquired by Frogel {\it et~al.} (1987) and Graham~{\it 
et~al.} (1987) (spectra published in Meikle {\it et al.} 1997). These 
were the first-ever NIR spectra of a Type~Ia event. The two spectra of 
Graham {\it et al.} are included in Fig~1. 
 
\subsection{SN~1991T} 
The photospheric-phase behaviour of SN~1991T was atypical. At peak it was 
exceptionally luminous. It also exhibited strong iron-group lines in its 
pre-maximum spectra, plus weaker-than-normal intermediate-mass spectral 
features. NIR spectra of SN~1991T and their analysis have been presented 
in Meikle {\it et al.} (1996) (day~$\sim$0), Spyromilio {\it et al.} 
(1994) (day~60), and Spyromilio {\it et al.} (1992) (day~338). The 
days~60 and 338 spectra are included in Figs.~1 \& 2 respectively. An 
optical spectrum at about the same phase as the day~60 NIR spectrum was 
obtained in AAT service time and is shown in Fig.~3. Also shown is an 
AAT service spectrum taken on day~403.  
 
\subsection{SN~1992G} 
The NIR spectrum of SN~1992G was obtained using the 75~line~mm$^{-1}$ 
grating and 58$\times$62 pixel array at three wavelength settings to 
cover the wavelength range 1.0-1.8~$\mu$m. A one pixel (3 arcsec) wide 
slit was used, oriented at a position angle of 21$^\circ$. The J-band 
was covered on day~39, and the H-band on day~40. Atmospheric OH lines 
were used to calibrate the wavelength scale. The flux standard was 
BS4078. No contemporary optical spectra were available.  However, in 
Fig.~3 we show a 17~day spectrum taken by R.~Lopez and G.~Gomez using 
the Faint Object Spectrograph (FOS) on the Isaac Newton Telescope (INT). 
We obtained the spectrum from the La Palma Data Archive in unreduced 
form. We reduced the spectrum using the Figaro data reduction package 
(Shortridge 1995).  CuAr or CuNe arc lamp spectra were used for 
wavelength calibration, and the spectrophotometric standard star, 
G138-31, provided flux calibration. The spectral coverage and resolution 
is given in Table~3. The optical spectrum suggests that SN~1992G was, 
spectroscopically at least, a normal Type~Ia supernova. 
 
\subsection{SN~1994ae}  
The day~175 NIR spectrum of SN~1994ae was acquired with the very 
recently upgraded CGS4 with a 256$\times$256 InSb array and the 
75~line~mm$^{-1}$ grating, under `shared risks' conditions. Owing to a 
technical problem, we had to use a 1.25 arcsec (1 pixel) wide slit, 
which meant that the absolute fluxing was less reliable than usual. The 
slit was placed east-west to minimize the gradient of galaxy background 
light.  The flux standard was HD~84800. The NIR spectrum is shown in 
Fig.~2. We also obtained an optical spectrum of SN~1994ae at 89~days 
using ISIS on the William Herschel Telescope (WHT). The flux standard 
was Feige 34. The optical spectrum is shown in Fig.~3 and is typical of 
a Type~Ia supernova at this phase.  Unfortunately we were unable to 
acquire an optical spectrum more closely contemporary with the IR 
observation. 
 
\subsection{SN~1995D} 
The day~92 NIR spectrum of SN~1995D was obtained with the same 
instrument configuration as for SN1994ae.  The data were reduced in the 
usual way (see above) using HD~77281 as the flux standard. A 
contemporary optical spectrum is shown in Fig~3. This 
spectrum is courtesy of A.~Filippenko and D.~Leonard and was taken on 
day +96 with the Lick Observatory Shane~3~m telescope using the cooled 
germanium spectrograph. The spectrum, which covers the 0.36 to 
1.02~$\mu$m range, is typical of a Type~Ia at this phase, showing an 
abundance of singly- and doubly-ionized Co and Fe emission lines. 
 
\subsection{SN~1995al} 
IR spectra were obtained at 94 and 176~days. For the day~94 spectrum the 
75 line mm$^{-1}$ grating was used.  In view of the good seeing that 
night, a 1-pixel-wide slit (1.25~arcsec) was selected, oriented 
east-west. The flux standard was HD~84800. Some difficulty in 
subtracting off the galaxy background was experienced owing to the 
steep and rapidly changing gradient in the supernova vicinity.  For the 
day~176 spectrum we used the 150~line~mm$^{-1}$ grating and a 
2-pixel-wide slit (2.5~arcsec) oriented 63$^\circ$~northwest to minimize 
the gradient of galaxy light along the slit. The flux standard used was 
HD 84800. We encountered a problem in the reduction of the 3rd-order 
J-band 0.985--1.30~$\mu$m spectrum, owing to inadequate blocking of 
2nd-order H-band emission. In order to remove this 2nd-order 
contamination the observed H-band spectrum was divided by the function 
for the blocking filter used in the H-band and multiplied by the 
blocking filter function for the J-band.  The wavelength of the 
resulting spectrum was then multiplied by 2/3 to produce an estimate of 
the J-band contamination spectrum.  This was then used to decontaminate 
the supernova and standard spectra.  The supernova spectrum was flux 
calibrated as normal using the decontaminated standard star spectrum. 
An optical spectrum (Fig.~3) was obtained at 149~days through INT 
service using the Intermediate Dispersion Spectrograph (IDS) and the 
R150V grating. As for the 176~day NIR observation, the slit was 
oriented at 69$^\circ$ to minimize the galaxy gradient.  Feige~19 was 
the flux standard. The optical spectrum is typical of a Type~Ia event 
at this phase. 
 
\subsection{SN~1996X}  
The day~19 NIR spectrum was acquired using the same CGS4 settings as for 
the day~176 SN~1995al spectrum. BS4935 was used as the flux standard. 
 
\subsection {Summary of the infrared spectra} 
All nine of the NIR spectra in the 13 to 176~day era extend to 
wavelengths as short as 1.1~$\mu$m and usually to 1.0~$\mu$m (Figs.~1 
\& 2). They {\it all} show a characteristically dramatic decline in the 
$\sim$1.02--1.13~$\mu$m region. In addition, spectra taken on or after 
day~40 usually exhibit a relatively isolated feature at 
$\sim$1.26~$\mu$m. All seven spectra taken between 13 and 94~days show a 
sharp rise at $\sim$1.50~$\mu$m. Thus, we confirm that the well-known 
red J-H colour results from a flux deficit between 1.13 and 1.50~$\mu$m 
and our spectra demonstrate that this deficit is responsible for the red 
colour to at least 94~days. Two of the spectra (91T at 60~d, 95D at 
92~d) span the 0.9 to 1.0~$\mu$m region. These reveal a strong, 
multi-peaked feature. In the range 1.5--1.8~$\mu$m, SN~1991T exhibits a 
relatively featureless band of emission on day~60, replaced by day~338 
by a prominent single peak at 1.65~$\mu$m.  Both SN~1995D and SN~1995al 
at $\sim$90~days show a distinct emission feature at 1.56~$\mu$m, with 
more complex blends in the 1.67--1.80~$\mu$m region. In the 
2--2.4~$\mu$m range between day~13 and 92 the spectra exhibit a very 
characteristic, and remarkably unchanging multi-peaked structure. 
  
\section {Spectral Model}
In order to identify and interpret the observed IR~features, we have
constructed a non-LTE nebular spectral model and applied it to all of the
spectra at epochs later than 90 days. For this initial investigation we
confined the model to lines of cobalt, iron and sulphur.  In the case of
a thermonuclear scenario, cobalt and iron are expected to dominate the
NIR and optical spectra. Strong lines of singly-- and doubly--ionized
sulphur are also expected to contribute to the multi-peaked structure in
the J--band. However, a complete description should also include other
lighter elements such as silicon and magnesium. 

Several simplifying assumptions are made. The electron temperature and
ionization state of the gas are constant throughout the nebula. 
In addition, the total line fluxes, including the effects of
line-trapping (see below) are calculated assuming a constant density.
Consequently the profiles of optically-thin lines are parabolic. In
general, a relatively weak continuum, probably due to blends of many
weak lines, was also present in the observed spectra.  This was
represented with a flat continuum in the model.  Since
cobalt and iron have similar ionization potentials in the singly-- and
doubly--ionized states, we set N$_{Co^+}$/N$_{Fe^+}$ and
N$_{Co^{2+}}$/N$_{Fe^{2+}}$ to be equal to the total cobalt/iron
abundance ratio.   The cobalt lines are
due to radioactive $^{56}$Co (the decay product of $^{56}$Ni) and the
iron lines come from both $^{56}$Fe (the stable decay product of
$^{56}$Co) and also from other stable iron isotopes formed before and
during the explosion. For these other iron isotopes, the mass is 19\%
relative to the total iron mass (Nomoto, Thielemann, and Yokoi 1984).

The $^{56}$Ni $\rightarrow$ $^{56}$Co $\rightarrow$ $^{56}$Fe light curve
hypothesis implies that we should expect the nebular-phase spectra to be
dominated by singly- and doubly-ionized lines of Fe and Co. We therefore
included model atoms (number of levels in brackets) for Fe$^+$ (44),
Fe$^{2+}$ (27), Co$^+$ (25) and Co$^{2+}$ (17). A 25-level Fe$^o$ atom
was also included since the red wing of one of its strongest lines,
[Fe~I]~1.443~$\mu$m, was covered by some of our data thus providing the
possibility of constraining the abundance of this species. We did not
include a Co$^o$ atom as no strong [Co~I] lines occur in our optical--NIR
spectral range. Transitions of [Fe~IV] and [Co~IV] do occur in our
spectral range, but they arise from energy levels with excitation
potentials higher than 32000 cm$^{-1}$ and 23000 cm$^{-1}$ respectively
and so are expected to be weak. Strong [S~II] and [S~III] lines exist in
the 0.95 and 1.02~$\mu$m region and so 5-level atoms of S$^+$ and
S$^{2+}$ were also included. However, since the major aim of this paper
is to test the $^{56}$Ni hypothesis, we have not included any other
intermediate-mass element transitions in either the NIR or optical
regions. 

During the supernova era covered by this work, for many of the lines 
the density is expected to fall below the critical value. Hence a 
non-LTE treatment is necessary i.e. we assume that the levels are 
populated by the combined effects of electron collisions in an 
isothermal gas and spontaneous decay. We include the effects of line 
trapping in an expanding atmosphere i.e. the downward radiative rates 
are modified by the Sobolev escape probability (Sobolev 1960). The 
detailed balance equations between thermal electron collisions and 
spontaneous decays are solved by matrix inversion. We have not treated 
the non-thermal excitation/de-excitation and ionization/recombination 
driven by the gamma-rays, high energy electrons and positrons. 
 
For [Fe~I], transition probabilities are taken from Fuhr, Martin \& Wiese 
(1988), and the collision strengths from Pelan \& Berrington (1996). For 
[Fe~II], transition probabilities are from Nussbaumer and Storey (1988b) 
\& Fuhr~{\it et~al.} (1988), and collision strengths from Pradhan \&
Zhang (1993 \& 1995).  For [Fe~III], transition probabilities are from
Nahar \& Pradhan (1996) and collision strengths from Zhang \& Pradhan
(1995) and Zhang (1996). For [Co~II], the transition probabilities are
from Nussbaumer \& Storey (1988a) and a compilation of Kurucz (1988). For
[Co~III], transition probabilities are from Fuhr {\it et al.} (1988),
Hansen (1984) and from a compilation of Kurucz (1988). For [S~II],
transition probabilities are from Mendoza \& Zappa (1982), and the
collision strengths from Cai \& Pradhan (1993) and the 1983 Mendoza
compilation. For [S~III], transition probabilities and collision
strengths are from the 1983 Mendoza compilation and the Kurucz list
(1975). A major uncertainty in this work is caused by the inaccurate
A-values for doubly--ionized cobalt
and the lack of collision strengths for singly-- and
doubly--ionized cobalt. Only for the a$^2$G--a$^4$F transitions are
collision strengths available (Kurucz 1988).  All the other collision
strengths for [Co~II] and [Co~III] were determined using the
approximation \[\Omega_{ij}=\omega g_i g_j\] where $\omega$=0.02 (Axelrod
1980; Graham, Wright \& Longmore 1987).  This has been found to yield
collision strengths with errors of up to about a factor of two relative
to precise calculations.

The models were reddened by the appropriate amount. For each case, the 
extinction was obtained by two methods.  In the first method, the 
supernova (B-V) colours were compared with the zero-extinction colour 
template of Phillips {\it et al.} (1992) or Doggett \& Branch (1985). In 
the second method, E(B-V) was found from the equivalent width (EW) of 
the interstellar Na~I~D absorption line, and applying the relation 
EW=4*E(B-V) (Barbon~{\it et~al.}~1990). It was found that these two 
methods produced values which usually agreed to within 20\%. Assuming 
E(B-V)=3.1A(V), the full spectral reddening function for each supernova 
was then obtained from the empirical extinction law of Cardelli {\it et 
al.} (1989). 
 
Estimation of the distances to the supernovae was a problem. An 
obvious method is to use the light curves of the actual supernovae ({\it 
cf} Reiss {\it et al.} 1995). However, a light curve was available for 
{\it only} SN~1991T. We therefore estimated the distances of the parent 
galaxies using the Aaronson {\it et al.} (1982) infall model (rms 
dispersion of 150kms$^{-1}$), taking the distance of the Virgo Cluster 
to be 16$\pm$1.5~Mpc (Van Den Bergh 1996). We note that for SN~1991T, 
this procedure gave a distance of 14.4~Mpc, in reasonable agreement 
with the light curve-based 12~Mpc quoted by Reiss {\it et al} (1995), 
and the 16.2$\pm$2~Mpc of H\"{o}flich (1995). The adopted distances are 
given in Table~4. 
 
While distance is clearly a critical parameter in absolute mass 
estimation it can, to a lesser extent, also influence the determination 
of the cobalt/iron mass ratio. As explained below, a critical step in the 
determination of the mass ratios is the matching of model line flux {\it 
ratios} to those observed. For the cases where the lines involved are 
optically thin and where the electron density is above the critical 
density, the line flux ratios are independent of distance. However, in 
situations of non-negligible optical depth or below critical density, the 
density can influence the line ratios. To match the flux in a particular 
line requires the population density, supernova radius and distance. The 
supernova radius is obtained from the observed expansion velocity and 
time since explosion.  For a given spectrum, a change in distance 
requires a density change to maintain the match to the data. However, the 
effects of optical depth and critical density mean that the required 
change in density for a given distance change may be different for 
different transitions.  Thus even when determining mass ratios it can 
be necessary to have an independent estimate of the distance although, 
as already indicated, the dependence of a given line {\it ratio} on 
distance/density is usually weak. 
 
Initially we set the Co/Fe mass ratio at the value expected assuming all 
the cobalt and iron results from $^{56}$Ni decay. Electron temperature 
and ion densities were then adjusted to reproduce the ratios of 
particular, prominent lines. (The total electron density was set by the 
ion densities). In addition to iron-group lines, [S~II]1.03 and 
[S~III]0.953~$\mu$m were included.  Adopting the same temperature as 
for the iron-group spectra, the abundance of S$^+$ and S$^{2+}$ was 
adjusted to match the strong features in the 0.9--1.1~$\mu$m region. To 
test the radioactive decay hypothesis, the Co/Fe mass ratio was then 
varied and the temperature and densities were adjusted where possible 
to provide a reasonable match to the spectra.  In this way we found the  
range of mass ratios which would provide a plausible fit to the data. 
Using the model fits, for each species we also determined the absolute 
mass assuming a radius given by the product of the expansion velocity
and the age of the supernova.

\section {Results} 
 
All the spectra with their best visually-matched models are shown in 
Figs.~4--12 and the model parameters are listed in Tables~4 and 5. 
Tables~6, 7 and 8 give the peak wavelengths of the main features present 
in each spectrum together with the identity and flux of the main 
contributing transitions derived from the models. It can be seen that 
there is a wealth of doubly-- and singly--ionized iron and cobalt lines 
in both the optical and infrared regions. In both the NIR and optical 
regions, for relatively isolated features of good S/N, the line widths 
were reasonably reproduced with expansion velocities in the range of 
8--8.5$\times$10$^3$~km/s. 
 
\subsection {Individual Supernovae} 
 
For SN~1995D at 92~d (Figs.~4, 5 \& 6) the model electron temperature and 
Co$^{2+}$ density were adjusted to provide the best visual match to the 
[Co~III]0.589/0.591~$\mu$m/ [Co~III]2.002~$\mu$m line flux ratio. (As noted 
above, the a$^4$F--a$^2$G [Co~III]0.589/0.591~$\mu$m transitions are two 
of the few cobalt transitions for which collision strengths {\it are} 
available.) The ionization ratio for cobalt was then determined by 
adjusting the relative abundance of Co$^+$ to match the 1.55~$\mu$m 
feature which is a blend dominated by [Co~II]1.547~$\mu$m and 
[Co~III]1.549~$\mu$m. The iron spectra were then formed by using the same 
electron temperature, assuming the $^{56}$Ni cobalt/iron mass ratio and 
adopting the same ionization ratio as for cobalt. In the NIR spectrum 
(Figs.~5 \& 6), it can be seen that in addition to the 1.55 and 
2.002~$\mu$m features, emission at 1.76, 2.15 and 2.24~$\mu$m are 
reproduced by [Co~III] and [Fe~III] transitions although the model flux
is somewhat weak here. However, emission in other parts of the NIR
spectrum remain unexplained.  These include regions around 1.00, 1.10,
1.27 and 1.65~$\mu$m.  While the emission at 1.27~$\mu$m is almost certainly 
due to a blend of [Co~III] and [Fe~II] lines, adjusting the temperature or
ionisation ratio to force a match here results in severe overproduction of
either [Co~III]0.589/0.591~$\mu$m or [Fe~II]0.44~$\mu$m.

For the optical spectrum (Fig.~4) the model is less successful. Many of
the discrepancies in this region may well be due to the inaccurate
A-values for doubly-ionised cobalt and for the [Fe~II]~a$^2$G term.
Another probable cause is the non-inclusion of intermediate-mass elements
in the model such as silicon and magnesium. A third possibility is
suggested by a preliminary calculation using a more complete model, which
shows that the ion fraction of Fe$^{2+}$/Fe$^+$ increases with radius.
The total flux which results is the average over this distribution and
may not be achievable from any single ionization ratio. A curious
discrepancy is that emission in the 0.43-0.44~$\mu$m region is
underproduced by the model, with the shortfall worsening with time. A
similar effect can be seen in the model fits of Ruiz-Lapuente (1992) for
SN~1986G at 257, 296 and 323 days. Using a more sophisticated model, Liu,
Jeffery \& Schultz (1997) also find this effect, although in their study
it only becomes a problem at about 1~year. They suggest that by these
late times a significant contribution to the Fe~II emission may come from
the recombination or charge transfer cascades from Fe$^{2+}$. They also
suggest time-dependent effects or emission from other species as
alternative explanations. Another interesting result from the model
comparison is that strong observed features at 0.47~$\mu$m and
0.589/0.591~$\mu$m are blueshifted with respect to the model
[Fe~III]~0.47~$\mu$m and [Co~III]~0.589/0.591~$\mu$m lines by
$\sim$3500~km/s and $\sim$1500~km/s respectively. Blueshifts of
$\sim$1000~km/s are also apparent for the features at 1.56 and
2.01~$\mu$m. We note that the 89~d optical spectrum of SN~1994ae exhibits
similar blueshifts to those seen in SN~1995D.

For SN~1995al at 94~d (Fig.~7) only an infrared spectrum was available. 
In this case we were unable to pin down the electron temperature, ion 
density or ionization ratio. Therefore we assumed the same electron 
temperature as for SN~1995D at a similar epoch. The Co$^+$+Co$^{2+}$ 
densities {\it and} the ionization ratios were adjusted to provide a 
match to the observed 
([Co~II]1.547+[Co~III]1.549~$\mu$m)/[Co~III]2.002~$\mu$m flux ratios. As 
with SN~1995D at 92~d, the 1.76~$\mu$m feature is partially reproduced by 
a blend of [Co~II] and [Co~III] lines.  Unfortunately the 2.15 and 
2.24~$\mu$m features were not covered by the observations.  Again, 
similarly to SN~1995D at 92~d, observed emission around 1.10 and 
1.65~$\mu$m is unaccounted for. However, the match to the feature at 
1.27~$\mu$m is more satisfactory.  The blueshifts seen in the NIR 
spectra of SN~1995D are not apparent in the SN~1995al data.  
 
For SN~1995al at +176d (Figs.~8 \& 9), the relatively low S/N and the 
smaller wavelength coverage in the NIR (Fig.~9) prompted us to use only 
the 149 day optical spectrum (Fig.~8), scaled to roughly match the epoch 
of the infrared spectrum, for determining the model parameters. The 
temperature was adjusted to provide a match to the observed
[Co~III]0.589/0.591~$\mu$m/[Fe~III]0.466/0.470~$\mu$m flux ratio. The
ionization ratio was then adjusted to match the
[Fe~II]0.716/0.717~$\mu$m flux.  The resulting NIR spectrum was then
compared with the data. Given the low S/N in the NIR, the model is
reasonably consistent with the observed NIR spectrum of SN~1995al.
Nevertheless, there is still unexplained emission around 1.26~$\mu$m
and 1.65~$\mu$m.  In the optical region, even allowing for the
additional deliberate model match in the 0.716~$\mu$m region, the match
is significantly better than for the $\sim$90d spectra.  In particular,
the wavelength agreement of the strong 0.47~$\mu$m and 0.59~$\mu$m
features with the model is much better than for SN~1995D at 92~d. There
is no evidence of a blueshift in the 0.59~$\mu$m feature, while at
0.47~$\mu$m the shift is less than $\sim$1000~km/s. Unfortunately we do
not have an earlier optical spectrum for SN~1995al to compare with the
$\sim$90~d SN~1995D and SN~1994ae spectra.
 
For SN~1994ae at 175~d (Fig.~10) only an infrared spectrum was available. 
Therefore a similar approach was adopted as for SN~1995al at 94d, using the 
same electron temperature as for SN~1995al at 176~d. The greater 
uncertainty in the model parameters due to the lack of an optical 
spectrum, was worsened by the very low S/N of the NIR spectrum, even 
after binning by a factor of four. The model appears to account for the 
emission at 1.03, 1.55 and 1.65~$\mu$m but there is unexplained emission 
around 1.10, 1.22 and, possibly, 1.75~$\mu$m. 

For SN~1991T a contemporary optical spectrum was derived by approximately
scaling the day~403 optical spectrum to the epoch of the day~338 NIR
spectrum (Figs.~11 \& 12) (see Spyromilio {\it et al.} 1992).  The model
electron temperature was adjusted to provide a match to the observed
[Fe~II]0.717~$\mu$m/[Fe~II]1.644~$\mu$m flux ratio. The ionization ratio
was adjusted to match the [Fe~III]0.466/0.470~$\mu$m flux. The cobalt
spectra were then formed by using the same model parameters, assuming the
radioactive scenario mass-ratio. The resulting IR/optical model,
dominated by iron line emission, is gratifyingly similar to the observed
spectra. Even the K-band emission seems to be reproduced by [Fe~III]
lines, although the S/N is low. As suggested by Ruiz--Lapuente (1992),
much of the unreproduced emission in the 0.7--0.95~$\mu$m region may be
due to (stable) $^{58}$Ni, via the transitions [Ni~II] 0.7380~$\mu$m and
[Ni~III] 0.7899~$\mu$m. In agreement with Spyromilio {\it et al.} (1992),
we conclude that the success of the model fit here supports the
prediction of the $^{56}$Ni scenario that over 96\% of the original
nickel should have decayed to iron by this epoch. We believe that the
remaining $\sim$4\% of cobalt is still apparent in the
[Co~III]~0.589/0.591~$\mu$m feature.

\subsection {Consistency Checks} 
 
Following Spyromilio {\it et al.} (1992), we applied two simple 
consistency checks to the individual models. Firstly, for each supernova 
we made an independent estimate of the atomic density by distributing 
1~M$_\odot$ of iron-group elements throughout a sphere defined by the 
maximum velocity and epoch. We found that the atomic densities derived 
in this way agreed to within a factor of three of the total densities 
obtained from the spectral models. 
 
Secondly, the heating and cooling rates of the gas were compared. Energy 
is deposited in the ejecta by the decay of $^{56}$Co at a rate of  
 
\[5.80\times10^{-13}e^{-t/113}(f+0.035)~erg s^{-1}atom^{-1}\] 
 
where t is the time in days after the explosion and f is the fraction of 
$\gamma$-rays absorbed (Axelrod 1980). f varies as t$^{-2}$ so that we 
would expect f-values of $\geq$0.5 at 3 months and $\sim$0.1 at 6~months 
to a year for the supernovae studied. We compared predicted heating 
rates with the three cases for which we have both optical and NIR 
spectra.  We find that the observed total optical+NIR luminosities 
correspond to cooling rates about a factor of two below the heating 
rates.  Given that, in each case, only part of the total supernova 
spectrum was used we conclude that the predicted heating rates were in 
reasonable agreement with the cooling rates. 
 
\subsection {Discussion of the model matches} 

We find that a spectral model based on $^{56}$Co-decay can be adjusted to
reproduce several of the prominent features in the NIR
and optical spectra of Type~Ia supernovae at three distinctly different
epochs (3, 6 and 11 months). At 3 and 6~months, feature identifications
include [Co~III]0.589, 0.591~$\mu$m, [Co~III]2.002~$\mu$m,
[Co~II]1.547~$\mu$m--[Co~III]1.549~$\mu$m, [Fe~III]0.466, 0.470~$\mu$m,
while at 11 months [Fe~II]0.717, [Fe~II]1.257 and [Fe~II]1.644~$\mu$m
are prominent. Plausible temperatures and densities are also found (cf.
Axelrod 1980).  There does exist some unexplained emission in the J--
and H--windows in the 3--6 month period.  However, the fact that this
does not occur at 1~year indicates that the problem may lie with the
modelling of the cobalt lines which contribute significantly at the
earlier epochs. We suspect that inaccuracies in the atomic data for 
cobalt may be at least part of the problem.

As pointed out earlier, there is a possibility that Na~I~D contributes
to the 0.59~$\mu$m feature. Since our wavelength coverage extends into
the NIR we are able to use Na~I 2.21~$\mu$m to estimate the strength of
Na~I~D.  For LTE conditions at a temperature of $\sim$7000~K we would
expect Na~I 2.21~$\mu$m to have a flux level of $\sim$3\% of Na~I~D. For
SN~1995D, the flux around 2.2~$\mu$m has a flux of about 1\% that of the
0.59~$\mu$m feature. While this does not rule out the presence of
Na~I~D in the 0.59~$\mu$m feature, it does seem unlikely that it could
be the dominant contributor.  We also note that in our model fitting we
found no need for the addition of sodium to improve the match.

While encouraging, these results alone do not rule out the possibility
that a similar success could be achieved with a completely different
model having the requisite number of freely-adjustable parameters. 
However, having adjusted our model to match the above specific
features, we find that a number of other spectral lines are also
successfully explained, particularly in the NIR. These are
[Co~III]1.742/1.764~$\mu$m, [Fe~III]2.146~$\mu$m and
[Fe~III]2.243~$\mu$m. The ability of our model to {\it predict} these
other features gives us added confidence in the appropriateness of the
$^{56}$Co-decay hypothesis.

In spite of the relative success of the $^{56}$Ni-based model in 
reproducing much of the observed NIR and optical nebular spectra, it is 
important to check the range of cobalt-to-iron mass ratios which 
uncertainties in the observed spectra, atomic data and distances would 
allow. At our earliest phase ($\sim$3~months), the electron density is 
above the critical value (i.e. the electrons are in LTE) and yet the 
lines have become optically thin.  Consequently, as explained above, 
derived mass ratios are distance independent.  However, A-value 
uncertainty is still important.  The mass ratio for a particular degree 
of ionization under LTE conditions can be expressed analytically as: 
 
\[\frac{M_{Co_n}}{M_{Fe_n}} = \frac{I_{ji}}{I_{j'i'}}\times  
\frac{\lambda_{ji}}{\lambda_{j'i'}}\times\frac{A_{j'i'}}{A_{ji}}\times  
\frac{g_{j'}}{g_{j}}\times\frac{e^{-Ej'/kT}}{e^{-Ej/kT}}\times  
\frac{Z_{Co_n}}{Z_{Fe_n}}\times\frac{m_{Co}}{m_{Fe}}\]  

where Co$_n$ and Fe$_n$ indicate ions of the same degrees of ionization,
I is the line flux, $\lambda$ the transition wavelength, A the Einstein
A-value, g the statistical weight, E the upper level energy, Z the
partition function, m the atomic mass and the subscripts ji and j'i'
represent the transitions in the relevant cobalt and iron ion lines
respectively.  At later epochs the spectrum is produced under
increasingly non--LTE conditions.  In general, this means that the
derived mass ratio is sensitive to uncertainties in both A-values and
collision strengths, as well as (to a lesser extent) distance, as
explained above. In this regime, we do not have a simple analytical
expression for the mass ratios.  For both LTE and non-LTE conditions, to
determine the effects of our imprecise knowledge of the atomic data we
assumed Gaussian distributions in their uncertainties and used a Monte
Carlo simulation to find the range of mass ratios for which a match to
the observed spectra was still obtainable.  A change in temperature will
also produce a change in the mass ratio.  The effect is strongest in the
Boltzmann factor, producing around $\pm$10\% in the mass ratio for
$\Delta$T=$\pm$500K, which is the typical temperature range that the
model fits permit. The partition functions are also sensitive to
temperature, but their {\it ratios} are much less sensitive, typically
remaining within $\pm$5\% for $\Delta$T=$\pm$1000K. In addition, distance
uncertainty can have a small influence on the mass ratio for non-LTE
conditions. Taking these various effects into account, we find that the
values of M$_{Co}$/M$_{Fe}$ are uncertain by factors of $\sim$1.5 to
$\sim$4 depending on the availability and quality of the data. These
ranges are shown in Table~9.

\section {
Absolute mass estimates} 
 
We also used the model fits to estimate the {\it absolute} masses of 
$^{56}$Ni produced in each supernova. In Table~10 we present the masses 
of the individual species, together with the total mass of $^{56}$Ni 
obtained by summing the singly- and doubly-ionized iron and cobalt. We 
include upper limits for neutral iron and cobalt.  The neutral iron limit 
was estimated from the region of the H-band which covers part of the 
[Fe~I]1.443~$\mu$m feature. The neutral cobalt limit 
was then derived assuming the same ionization fraction and the radioactive 
decay mass ratio. We find that the contribution of neutral cobalt and 
iron to the total iron-group mass is generally less than 15\%. In the case 
of SN~1995al at +176 days the H-band spectrum does not extend far enough 
blueward to allow us to place an upper limit on neutral iron and cobalt. 
Unfortunately, triply-ionized cobalt or iron do not produce any prominent 
emission lines in the NIR or optical region and so a significant mass of 
either cannot be ruled out by the data. Axelrod (1980) estimates that the 
contribution of triply-ionized material could be about 10\% of the total 
mass at 3~months rising to $\sim$25\% at 6~months and $\sim$40\% at one 
year. These estimates are therefore also included in Table~10. However, 
the values for the total $^{56}$Ni mass shown in Table~10 are 
conservative as they are obtained by summing the singly- and 
doubly-ionized iron and cobalt only. 

As with the mass ratio study, absolute mass determination is subject to
atomic data uncertainty. In general, however, it is more sensitive to
temperature, distance and extinction uncertainty. The effect of all these
uncertainties is that the derived masses are uncertain by factors of
$\sim$1.5 to $\sim$4, again depending on the availability and quality of
the data (see Table 10). We note that the level of uncertainty in the
mass {\it ratio} estimates is not significantly less than this, in spite
of their lower sensitivity to distance and temperature errors. This is
because the atomic data uncertainties, dominate the overall error for
both the relative and absolute masses.

For SNe~1991T, 1995D and 1995al our estimates for the total $^{56}$Ni mass 
of all four ionization states are in the range 0.37--0.90~M$_{\odot}$. 
If we confine our 
mass estimate to the directly measured singly- and doubly-ionized 
species only, the mass is 0.27--0.67M$_\odot$. 
The total $^{56}$Ni mass for SN~1994ae is found to be 
only about 0.2~M$_{\odot}$ but it must be remembered that
the data in this case are of very low signal-to-noise.
and at least twice this amount is possible.  
Thus, in spite of the uncertainties, the values found point towards the 
ejection of substantial masses of $^{56}$Ni, and so support the basic 
thermonuclear scenario.  
 
\section {Summary and Conclusions} 
In this paper we have presented the most extensive NIR spectral coverage 
hitherto of Type~Ia supernovae, spanning 13 to 338~days after maximum 
blue light. In some cases, contemporary or near-contemporary optical 
spectra were also shown. The NIR spectra are highly structured 
throughout the J-, H- and K-bands. In particular, all nine spectra in 
the 13 to 176~day era show a characteristically steep decline in the 
$\sim$1.02--1.13~$\mu$m region, while all seven spectra taken between 
13 and 94~days show a sharp rise at $\sim$1.50~$\mu$m. We conclude that 
the 1.13--1.50~$\mu$m deficit is responsible for the red colour for the 
entire period from 13 to at least 94~days. 
 
Comparison of the nebular-phase spectra with a simple non-LTE spectral 
model based on $^{56}$Ni-decay abundances, indicates that significant 
fractions of both the NIR and optical spectra are made up of iron and 
cobalt lines. Singly-ionized sulphur also contributes strongly to the 
emission around 1~$\mu$m. We conclude that the 1.13-1.50~$\mu$m deficit is 
due to a lack of emission lines from species which dominate the rest of 
the IR/optical spectrum, as originally proposed by Spyromilio, Pinto
and Eastman (1994). Identified NIR features include
[Fe~II]1.257~$\mu$m, [Fe~I]1.443~$\mu$m,
[Co~II]1.547~$\mu$m--[Co~III]1.549~$\mu$m, [Fe~II]1.644~$\mu$m,
[Co~III]1.742--1.764~$\mu$m, [Co~III]2.002~$\mu$m, [Fe~III]2.146~$\mu$m
and [Fe~III]2.243~$\mu$m.  Failure to account for some of the NIR
emission in the 3--6 month period may be partly due to inadequate
atomic data for cobalt. 

The model comparison for SN~1995D at 92~d revealed blueshifts in the
features at 0.47~$\mu$m, 0.589/0.591~$\mu$m, 1.56~$\mu$m and
2.01~$\mu$m.  Similar shifts were seen in the optical spectrum of
SN~1994ae at 89~d.  However, for the 94~d NIR spectrum and 149~d optical
spectrum of SN~1995al, the blueshifts were much smaller or
non-existent.  At later epochs the blueshifts diminished or disappeared
altogether. The fact that the SN~1995D blueshifts occurred at several
places over the entire NIR-optical spectrum tends to argue against
blending with unidentified lines as the general cause, although this may
be a partial explanation for the larger blueshift at 0.47~$\mu$m. In addition, 
all the model lines are optically thin and so the effect cannot be simply one
of optical depth in a homogeneous medium. We therefore suggest the
ejecta are actually inhomogeneous and that the blueshifts are due to
density enhancements which are optically thick. Some support for this
is provided by the fact that small or no blueshifts were seen at 6 and
11~months.
 
Our $^{56}$Ni-based model, using plausible temperatures and densities, 
both accounts for and predicts features in the NIR and optical regions. 
In particular it shows that the cobalt lines decay with time relative 
to the iron lines and that this is due to the decline in the abundance 
of cobalt at the $^{56}$Co-decay rate.  This gives us confidence in the 
appropriateness of the $^{56}$Ni-decay light-curve hypothesis. 
Nevertheless, there are important discrepancies between the model 
predictions and the observations.  This is probably due to a 
combination of unidentified lines, inaccurate atomic data, and the use  
of a simple spectral model.  In spite of 
the large uncertainties in various parameters, {\it we nevertheless 
conclude that the mass of ejected $^{56}$Ni is large.}  We believe that, 
on balance, our results support the basic thermonuclear explosion 
scenario and argue against the core-collapse model of Colgate {\it et 
al.} (1997), at least for most Type~Ia supernovae. The range of values 
encompasses predictions of white-dwarf-merger, delayed-detonation and 
sub-Chandrasekhar mass models. 
 
The work presented here is only a first attempt to describe and 
interpret the late-time NIR behaviour of Type~Ia supernovae.  Compared 
with the optical band, the quantity and quality of spectra are still 
much inferior.  However, the introduction of more sensitive NIR detectors 
coupled with adaptive optics means that rapid improvement in data 
quality can be anticipated in the near future. Indeed it may become 
possible to detect the presence of other radioactive isotopes such as 
$^{57}$Co ({\it cf} Varani et al. 1990). The development and 
application of more self-consistent and comprehensive spectral 
synthesis codes such as EDDINGTON (Eastman \& Pinto 1993) coupled with 
our steadily improving knowledge of the distance scale out to 
$\sim$50~Mpc will lead to elucidation of the appropriate thermonuclear 
explosion model.  However, this will not be possible if better atomic 
data does not become available.  This is the key improvement required 
for the field to progress. \\ 
%\newline 
 
\bf{Acknowledgements} \\ 
\normalsize 
\\ 
We would especially like to thank Ron Eastman for helpful discussions 
regarding the spectral modelling. We are indebted to Alexei Filippenko 
for providing us with a contemporary optical spectrum of SN~1995D. Our 
thanks go to Jason Spyromilio for assisting with the the SN~1992G 
spectrum, for providing us with the +403 day optical spectrum of 
SN~1991T, and for many constructive comments in his r\^{o}le as referee.
We much appreciate the staff at UKIRT for their excellent 
support.  We thank John Pilkington of the RGO for measuring astrometric 
positions. The United Kingdom Infrared Telescope is operated by the Joint 
Astronomy Centre on behalf of the U.K. Particle Physics and Astronomy 
Research Council (PPARC). The Isaac Newton and William Herschel 
Telescopes are operated on the island of La Palma by the Royal Greenwich 
Observatory in the Spanish Observatorio del Roque de los Muchachos of the 
Instituto de Astrofisica de Canarias. EJCB is supported by a PPARC 
studentship. 
\newpage 
\noindent 
\bf{REFERENCES} 
\normalsize 
 
\noindent Aaronson M., Huchra J., Mould J., Schechter P.L., Tully R.B., 1982, ApJ, 258, 64 \\ 
\\ 
\noindent Axelrod T.S., 1980a, Ph.D. thesis, UCRL-52994, University of 
California, Santa Cruz \\ 
\\ 
\noindent Axlerod T.S., 1980b, in J.C. Wheeler ed, Type I Supernovae,  
University of Texas, Austin, p80 \\ 
\\ 
\noindent Barbon R., Cappellaro E., Turatto M., 1984, A\&A, 135, 27 \\ 
\\ 
\noindent Barbon R., Benetti S., Rosino L., Cappellaro E., Turatto, M., 
1990, A\&A, 237, 79 \\ 
\\ 
\noindent Branch D., 1980, in Meyerott R., Gillespie G.H., eds,  
Supernovae Spectra, American Institute of Physics, New York, p39  \\ 
\\ 
\noindent Cai W. \& Pradhan A.K., 1993, ApJ Suppl., 88, 329 \\  
\\ 
\noindent Cardelli J., Clayton G., Mathis J., 1989, ApJ, 345, 245 \\ 
\\ 
\noindent Colgate S.A., McKee C., 1969, ApJ, 157, 623 \\ 
\\ 
\noindent Colgate S.A., 1997, in Ruiz-Lapuente R., Canal R., Isern J., eds,  
Thermonuclear Supernovae, NATO ASI Series, p273   \\ 
\\ 
\noindent Daly P.N., 1996, SUN/27, Starlink Project, CLRC  \\ 
\\ 
\noindent Doggett J.B. \& Branch D., 1985, ApJ, 90, 2303 \\ 
\\ 
\noindent Eastman R.G. \& Pinto P.A., 1993, ApJ, 412, 731 \\ 
\\ 
\noindent Elias J.H., Frogel J.A., Hackwell J.A., Persson S.E., 1981, 
ApJ, 251, L13 \\ 
\\ 
\noindent Elias J.H., Matthews K., Neugebauer G., Persson S.R., 1985, 
ApJ, 296, 379 \\ 
\\ 
\noindent Evans R., 1986, IAU Circ. 4208 \\ 
\\ 
\noindent Fairall A., 1986, IAU Circ. 4210 \\ 
\\ 
\noindent Filippenko A.V., Richmond M.W., Matheson T., Shields J.C., 
Burbidge E.M., Cohen R.D., Dickinson M., Malkan M.A., Nelson B., Pietz 
J., Schlegel D., Schmeer P., Spinrad H., Steidel C.C., Tran H.D., Wren W., 
1992 ApJ, 384, L15 \\ 
\\ 
\noindent Frogel J.A., Gregory B., Kawara K., Laney D., Phillips M.M., 
Terndrup D., Vrba F., Whitford A.E., 1987, ApJ, 315, L129 \\ 
\\ 
\noindent Fuhr J.R., Martin G.A., Wiese W.L., 1988, Atomic Transition 
Probabilities, New York \\ 
\\ 
\noindent Graham J.R., 1986, MNRAS, 220, 27p \\ 
\\ 
\noindent Graham J.R., Baas F., Geballe T.R., Smith M.G., Longmore A.J., 
Williams P.M., 1987, preprint \\ 
\\ 
\noindent Graham J.R., Wright G.S., Longmore A.J., 1987, ApJ, 313, 847 \\ 
\\ 
\noindent Hansen, 1984, ApJ, 277, 435 \\ 
\\ 
\noindent Heathcote D., Cowley A., Hutchings J., 1986, IAU Circ. 4210 \\ 
\\ 
\noindent Hill P.W., 1986, IAU Circ. 4210 \\ 
\\ 
\noindent H\"{o}flich P., Khokhlov A.M., Wheeler J.C., 1995, ApJ, 444, 831 \\ 
\\ 
\noindent H\"{o}flich P., Khokhlov A.M., 1996, ApJ, 457, 500 \\ 
\\ 
\noindent Hoyle F., Fowler W.A., 1960, ApJ, 132, 565 \\ 
\\ 
\noindent Hummer D.G., Berrington K.A., Eissner W., Pradhan A.K., Saraph
H.E., Tully J.A., 1993, A\&A, 279, 298 \\
\\
\noindent Iben I. \& Tutukov A., 1984, ApJ Suppl, 55, 335 \\ 
\\ 
\noindent Jeffery D., Leibundgut B., Kirshner R.P., Benetti S., Branch 
D., Sonneborn G., 1992, ApJ, 397, 304 \\ 
\\ 
\noindent Khokhlov A.M., 1991a, A\&A, 245, 114 \\ 
\\ 
\noindent Khokhlov A.M., 1991b, A\&A, 245, L25 \\ 
\\ 
\noindent Kirshner R.P., Oke J.B., Penston M., Searle L., 1973, ApJ, 185, 
303 \\ 
\\ 
\noindent Kirshner R.P. \& Kwan J., 1975, ApJ, 197, 415 \\ 
\\ 
\noindent Kirshner R.P., 1986, IAU Circ. 4216 \\ 
\\ 
\noindent Kirshner R.P., Jeffery D.J., Leibundgut B., Challis P.M., 
Sonneborn G., Phillips M.M., Suntzeff N.B., Smith R.C., Winkler P.F., 
Winge C., Hamuy M., Hunter D.A., Roth K.C., Blades J.C., Branch D., 
Chevalier R.A., Fransson C., Panagia N., Wagoner R.V., Wheeler J.C., 
Harkness R.P., 1993, ApJ, 415, 589 \\  
\\ 
\noindent Kuchner M.J., Kirshner R.P., Pinto P.A., Leibundgut B., 1994, 
ApJ, 426, L89 \\ 
\\ 
\noindent Kurucz R.L. \& Peytremann E., 1975, SAO Special Report 362 \\ 
\\ 
\noindent Kurucz R.L., 1988, in McNally M., ed, Trans. IAU, XXB, Dordrecht:  
Kluwer, 168-172 \\ 
\\ 
\noindent Livne E., 1990, ApJ, 354, L53 \\ 
\\ 
\noindent Livne E., Glasner A.S., 1990, ApJ, 361, 244 \\ 
\\ 
\noindent Livne E. \& Arnett D., 1995, ApJ, 452, 62 \\ 
\\ 
\noindent Lynch D.K., Rudy R.J., Rossano G.S., Erwin P., Puetter R.C., 
1990, AJ, 100, 223 \\ 
\\ 
\noindent Lynch D.K., Erwin P., Rudy R.J., Rossano G.S., Puetter R.C., 
1992, AJ, 104, 1156 \\ 
\\ 
\noindent Mazzali P.A., Danziger I.J., Turatto M., 1995, A\&A, 297, 509 \\ 
\\ 
\noindent Meikle W.P.S., Bowers E.J.C., Pinto P.A., Eastman R.G., Cumming 
R.J., Geballe T.R., Lewis J.R., Walton N.A., 1997, in Canal R., 
Ruiz-Lapuente P., Isern J., eds, Proc. NATO Advanced Study Institute on 
Thermonuclear Supernovae, Kluwer, The Netherlands, p53  \\ 
\\ 
\noindent Meikle W.P.S., Cumming R.J., Geballe T.R., Lewis J.R., Walton 
N.A., Balcells M., Cimatti A., Croom S.M., Dhillon V.S., Economou F., 
Jenkins C.R., Knapen J.H., Lucey J.R., Meadows V.S., Morris P.W., 
P\'{e}rez-Fournon I., Shanks T., Smith L.J., Tanvir N.R., Veilleux S., 
Vilchez J., Wall J.V., 1996, MNRAS, 281, 263 \\ 
\\ 
\noindent Mendoza C. \& Zeippen C.J., 1982, MNRAS, 198, 127 \\ 
\\ 
\noindent Mendoza C., 1983, in Proc. Symposium on Planetary Nebulae, 
Dordrecht, London, England, p143-172 \\ 
\\ 
\noindent Meyerott R.E., 1980, ApJ, 239, 257 \\ 
\\ 
\noindent Meurer G., 1986, AUC Circ. 4216 \\ 
\\ 
\noindent Nahar S.N. \& Pradhan A.K., 1996, A\&A, 119, 509 \\ 
\\ 
\noindent Nomoto K., Thielemann F-K., Yokoi K., 1984, ApJ, 286, 644 \\ 
\\ 
\noindent Nugent P., Baron E., Hauschildt P.H., Branch D., 1994, BAAS, 
185, 7902 \\  
\\ 
\noindent Nussbaumer H. \& Storey P.J., 1988a, A\&A, 200, L25 \\ 
\\ 
\noindent Nussbaumer H. \& Storey P.J., 1988b, A\&A, 193, 327 \\ 
\\ 
\noindent Pankey T., 1962, Ph.D. thesis, Howard University \\ 
\\ 
\noindent Pelan J. \& Berrington K.A., 1996, preprint \\ 
\\ 
\noindent Phillips M. \& Geisler D., 1986, IAU Circ. 4210 \\ 
\\ 
\noindent Phillips M. \& Hamuy M., 1991, IAU Circ. 5239 \\  
\\ 
\noindent Phillips M.M., Phillips A.C., Heathcote S.R., Blanco V.M., 
Geisler D., Hamilton D., Suntzeff N.B., Jablonski F.J., Steiner J.E., 
Gowley A.P., 1987, PASP, 99, 592 \\ 
\\ 
\noindent Phillips M.M., Wells L.A., Suntzeff N.B., Hamuy M., Leibundgut 
B., Kirshner R.P., Foltz C.B., 1992, AJ, 103, 1632 \\  
\\ 
\noindent Phillips M.M., 1993, ApJ, 413, L105 \\ 
\\ 
\noindent Pradhan A.K. \& Zhang H.L., 1993, ApJ, 409, L77 \\ 
\\ 
\noindent Reiss A.G., Press W.H., Kirshner R.P., 1995, ApJ, 438, L17 \\ 
\\ 
\noindent Ruiz-Lapuente P., 1992, Ph.D. thesis, University of Barcelona \\ 
\\ 
\noindent Schlegel E.M., 1995, ApJ, 109, 2620 \\
\\
\noindent Shortridge K., 1995, Starlink Project, CLRC \\ 
\\ 
\noindent Sobolev V.V., 1960, Moving Envelopes of Stars, Harvard University \\ 
\\ 
\noindent Spyromilio J., Meikle W.P.S., Allen D.A., Graham J.R., 1992, 
MNRAS, 258, 53p \\ 
\\ 
\noindent Spyromilio J., Pinto P.A., Eastman R.G., 1994, MNRAS, 266, L17 \\ 
\\ 
\noindent Thielemann F-K., Nomoto., Yokoi., 1986, A\& A, 158, 17 \\ 
\\ 
\noindent Tonry J. \& Strauss M., 1986, IAU Circ. 4210 \\ 
\\ 
\noindent Van Den Bergh S., 1996, PASP, 108, 1091 \\ 
\\ 
\noindent Varani G-F., Meikle, W.P.S., Spyromilio J., Allen D.A., 1990, 
MNRAS, 245, 570 \\ 
\\ 
\noindent Wamsteker W., Gilmozzi R., Gry C., Machetto F., Panagia N., 
1986, IAU Circ. 4216 \\ 
\\ 
\noindent Weaver J.C., Swartz D.A., \& Harkness R.P., 1993, Phys. Rep., 
227, 113 \\ 
\\ 
\noindent Wells L.A., Phillips M.M., Suntzeff N.B., Heathcote S.R., Hamuy 
M., Navarrete M., Fernandez M., Weller W.G., Schammer R.A., 1994, AJ, 
108, 2234 \\ 
\\ 
\noindent Woosley S.E., Weaver T.A., 1994a, in Bludman S., Mochkovitch 
R., Zinn-Justin J., eds, Proc. Les Houches Session LIX, Elsevier Science, 
Paris, p.63 \\ 
\\ 
\noindent Woosley S.E. \& Weaver T.A., 1994b, ApJ, 423, 371 \\ 
\\ 
\noindent Zhang H.L. \& Pradhan A.K., 1995, ApJ, 293, 953 \\ 
\\ 
\noindent Zhang H.L. \& Pradhan A.K., 1995, J.Phys.B: At. Mol. Opt. 
Phys., 28, 3403 \\ 
\\ 
\noindent Zhang H.L., 1996, A\&A, 119, 523 \\ 
\normalsize
\newpage 
\begin{table*} 
\footnotesize 
\centering 
\begin{minipage}{160mm} 
\caption[]{Type Ia supernovae observed spectroscopically in the infrared} 
\centering 
\hspace{1cm} 
\begin{tabular}{ccccccccc} \hline \hline 
 
SN & Parent & \multicolumn{2}{c}{Position (2000.0)} & Offset from & Discovery  
& Vmax & t$_{Bmax}$  & Discovery \\ 
   & Galaxy &          RA    &      Dec             & Gal. nucl. & Date     &        &  & IAU    \\  
   &        & (hr min sec)  & (deg amin asec)           & (asec) &                &        &  & Circular  \\ \hline   
\\ 
1986G & NGC 5128 & 13 25 36.55 & -43 01 51.7      & 120E 60S & 3-05-86  
& 11.4 & 11-05-86 & 4208 \\ 
1991T & NGC 4527 & 12 34 10.18  & +02 39 56.79    & 26E 45N    & 13-04-91 
& 11.5 & 28-04-91 & 5239     \\ 
1992G & NGC 3294 & 10 36 16.52 & +37 19 13.7      & 27E 10.5S  & 
9-02-92 & 13.7  & 20-02-92  & 5452  \\ 
1994ae & NGC 3370 & 10 47 01.99 & +17 16 32.1     & 30.3W 6.1N & 
14-11-94 & 12.5 & 27-11-94   &  6105    \\ 
1995D & NGC 2962 & 09 40 54.80 & +05 08 26.7      & 11E 90.5S & 10-2-94 
& 13.4 & 17-02-95    & 6134     \\ 
1995al & NGC 3021 & 09 53 53.02 & +33 18 58.7     & 15.0W 2.9S & 
1-11-95 & 13.0 & 30-10-95    & 6255     \\ 
1996X & NGC 5061 & 13 20 46.44 & -27 06 28.1      & 52W 31S & 12-4-96 & 
13.2 & 13-04-96  & 6380           \\  
\\ \hline 
\normalsize 
\end{tabular} 
\end{minipage} 
\end{table*} 

\begin{table*} 
\centering 
\begin{minipage}{160mm} 
\caption[]{Log of infrared spectroscopy} 
\centering 
\vspace{5mm} 
\begin{tabular}{cccccc} \hline \hline 
\multicolumn{1}{c}{Source} & Date UT & Epoch & Telescope/Instrument &  
$\lambda\lambda$ & Resolution  \\  
&& \multicolumn{1}{c}{(d)} 
&& \multicolumn{1}{c}{($\mu$m)} 
& \multicolumn{1}{c}{(kms$^{-1}$)} \\ \hline 
\\ 
SN1986G & 1986 May 24 & +13 & UKIRT/CGS2 & 1.056--1.822 & ~500 \\ 
&&&&                  1.935--2.487 & ~500 \footnote{The J, H and K  
observations which make up most of the `day +13' 
spectrum were actually obtained on days +12, +13 and +14 respectively. 
Smaller fill-in sections were obtained on day +18 (1.767--1.822~$\mu$m, 
1.935--1.971~$\mu$m) and day +22 (1.361--1.484~$\mu$m). The 
1.361--1.484~$\mu$m section was taken with a CVF.} \\ 
& 1986 June 10& +30 & UKIRT/CGS2 & 1.079--1.356 & ~500 \\ 
&& +31 &&                  1.945--2.517 & ~500 \footnote{In the `day +30'  
spectrum the 1.079-1.356~$\mu$m section was 
taken on day +30, while the 1.945--2.517~$\mu$m was taken on day +31.} \\ 
SN1991T & 1991 June 26 & +60 & AAT/IRIS & 0.90--1.31 & 500  \\ 
&&&& 1.41--1.80 & 800 \\ 
&&&& 1.91--2.45 & 550 \\ 
& 1992 Mar 31  & +338 & UKIRT/CGS4 & 1.18--1.36 & 800 \\ 
&&&& 1.42--1.8 & 800 \\ 
&&&& 2.07--2.46 & 800 \\ 
SN1992G & 1992 Mar 30  & +39 & UKIRT/CGS4 & 1.0--1.8 & 600 \\ 
SN1994ae & 1995 May 20  & +175 & UKIRT/CGS4 & 1.0--1.35 & 200 \\ 
& 1995 May 19 & +174 && 1.43--2.09 & 240 \\ 
SN 1995D & 1995 May 20  & +92 & UKIRT/CGS4 & 1.015--1.224 & 350 \\ 
& 1995 May 19 & +91 && 1.43--2.09 & 400 \\ 
& 1995 May 21 & +93 && 1.94--2.50  & 350 \\ 
SN1995al & 1996 Feb 6 & +94 & UKIRT/CGS4 & 1.0--1.34 & 250 \\ 
&&&& 1.42--2.08 & 600 \\ 
& 1996 Apr 28 & +176 & UKIRT/CGS4 & 0.985--1.30 & 1000 \footnote{The  
J spectra were made up of observations taken on days +175, 
+176 and +177.} \\ 
&&&& 1.49--1.80 & 400 \footnote{The H spectra were made up of observations  
taken on days +175 
and +176.} \\ 
& 1996 Apr 29 & +177 && 2.135--2.470 & 350 \\ 
SN1996X & 1996 Apr 29 & +19 & UKIRT/CGS4 & 0.995--1.30 & 300 \\  
&&&& 1.445--1.77 & 300 \\ 
&&&& 1.97--2.138 & 200 \\  
\\ \hline 
\end{tabular} 
\end{minipage} 
\end{table*} 
\begin{table*} 
\centering 
\begin{minipage}{160mm} 
\caption[]{Log of optical spectroscopy} 
\centering 
\vspace{5mm} 
\begin{tabular}{lccccc} \hline \hline 
\multicolumn{1}{c}{Source} & Date UT & Epoch & Telescope/Instrument  
                & $\lambda\lambda$  & Resolution  \\ 
&& \multicolumn{1}{c}{(d)} 
&& \multicolumn{1}{c}{($\mu$m)} 
& \multicolumn{1}{c}{(kms$^{-1}$)} \\ \hline 
\\ 
SN1991T & 1991 June 26 & +60 & AAT/Cass. spectrograph  & 0.38--0.95 & 900   \\ 
SN1991T & 1992 June 4 &  +403 & AAT/Cass. spectrograph  & 
0.32--1.00 & 900 \\ 
SN1992G & 1992 Mar 8 & +17 & INT/FOS  & 0.37--1.00 & 600 (1st ord) \\ 
&&&&& 800 (2nd ord) \\ 
SN1994ae & 1995 Feb 23 & +89 & WHT/ISIS & 0.34--0.92 & 300 \\ 
SN1995D & 1995 May 24 & +96 & Lick/Shane 3-m  & 0.31--1.04 & 200   \\ 
SN1995al & 1996 Apr 1  & +149 & INT/IDS   & 0.365--0.90 & 600 \\ 
\\ 
\hline 
\end{tabular} 
\end{minipage} 
\end{table*} 
\begin{table*} 
\centering 
\begin{minipage}{160mm} 
\caption[]{Fixed Parameters for the Spectral Synthesis Model Fits} 
\centering 
\vspace{5mm} 
\begin{tabular}{lccccc} \hline \hline 
\multicolumn{1}{c}{Source} & Epoch & D & Z & Av & N$_{Co^+}$/ \\ 
& \multicolumn{1}{c}{(d)} & 
\multicolumn{1}{c}{(Mpc)} & & &  
\multicolumn{1}{c}{N$_{Fe^+}$} \\ \hline 
\\ 
SN 1995D & 92 & 27.0$\pm$3.2 & 1.0065 & 0.6$\pm$0.1 & 0.66 \\ 
SN 1995al & 94 & 23.8$\pm$2.9 & 1.005 & 0.8$\pm$0.1 & 0.66 \\ 
          & 174 & & & & 0.25 \\ 
SN 1994ae  & 173 & 21.8$\pm$2.6 & 1.004 & 0.2$\pm$0.2 & 0.25 \\ 
SN 1991T & 338 & 14.4$\pm$1.7 & 1.0058  & 0.66$\pm$0.23 & 0.04 \\   
\\ \hline 
 
\end{tabular} 
\end{minipage} 
\end{table*} 

\begin{table*} 
\centering 
\footnotesize 
\begin{minipage}{160mm} 
\caption[]{Variable Parameters for the Spectral Synthesis Model Fits} 
\centering 
\vspace{5mm} 
\begin{tabular}{lcccccccc} \hline \hline 
\multicolumn{1}{c}{Source} & T$_e$ & 
N$_{Fe+Co}$\footnote{Total iron and cobalt ion density} &  neutral & singly & doubly &  
N$_S$\footnote{Total sulphur ion density} & N$_{S^+}$/ &
V$_{exp}$\footnote{Expansion velocity of supernova ejecta} \\ 
& \multicolumn{1}{c}{(K)} 
& \multicolumn{1}{c}{(cm$^{-3}$)} 
& \multicolumn{1}{c}{atom} 
& \multicolumn{1}{c}{ionized} 
& \multicolumn{1}{c}{ionized} 
& \multicolumn{1}{c}{(cm$^{-3}$)} 
& \multicolumn{1}{c}{N$_{S^{2+}}$} 
& \multicolumn{1}{c}{($\times$10$^3$kms$^{-1}$)} \\ 
& & & \multicolumn{1}{c}{fraction} 
& \multicolumn{1}{c}{fraction} 
& \multicolumn{1}{c}{fraction} 
& &  & \\ \hline 
\\ 
SN 1995D & 7000 & 3.9e6 & 0.2 & 0.2 & 0.6 & 7e5 & 0.5 & 8.0 \\ 
Figs~4, 5 \& 6 & & & & & & & & \\ 
SN 1995al at 94d & 7000 & 6.5e6 & 0.1 & 0.1 & 0.8  & 7e5 & 0.5 & 8.5 \\ 
Fig~7 & & & & & & & & \\ 
SN 1995al at 176d & 6000 & 5.8e5 & --- & 0.1 & 0.9 & 1.5e5 & 0.5 & 8.5  \\ 
Figs~8 \& 9 & & & & & & & & \\ 
SN 1994ae  & 6000 & 3e5 & 0.1 & 0.2 & 0.7 & 8.8e4 & 0.5 & 8.5 \\ 
Fig~10 & & & & & & & & \\ 
SN 1991T & 5600 & 1.7e5 & 0.01 & 0.15 & 0.84 & --- & --- & 8.5 \\ 
Figs~11 \& 12 & & & & & & & & \\ 
\\ 
\hline 
\end{tabular} 
\end{minipage} 
\normalsize 
\end{table*} 
\clearpage 
\begin{table*} 
\centering 
\footnotesize 
\begin{minipage}{160mm} 
\caption[]{Modelled Optical Emission Features of SNe Spectra} 
\centering 
\hspace{1cm} 
\begin{tabular}{clcccccc} \hline \hline 
& &  
\multicolumn{2}{c} {SN 1995D} &  
\multicolumn{2}{c} {SN 1995al} & 
\multicolumn{2}{c} {SN 1991T} \\  
& &  
\multicolumn{2}{c} {+92d} &  
\multicolumn{2}{c} {+176d} & 
\multicolumn{2}{c} {+363d}  \\  
$\lambda_{rest}^a$ & 
\multicolumn{1}{c} {Identification} & 
Line Flux$^b$ & $\lambda_{peak}\footnote{$\lambda$ is in microns}$ & 
Line Flux\footnote{Line flux is $\times$ 10$^{-15}$erg s$^{-1}$
cm$^{-2}$}  
& $\lambda_{peak}^a$ & Line Flux$^b$ & $\lambda_{peak}^a$ \\ 
\hline 
\\ 
0.4069\footnote{Blend of two lines} & [SII] 3p$^3$ $^2$P$_{3/2}$--3p$^3$  
$^4$S$_{3/2}$  & --- & --- & 13.4 & 
0.4095 & --- & --- \\  
0.4078$^c$ & [S II] 3p$^3$ $^2P_{1/2}$--3p$^3$ 
$^4S_{3/2}$ & --- & --- & 3.43 & & --- & --- \\  
0.4251\footnote{Blend of eight lines} & [Fe II] 
a$^4F_{5/2}$--b$^2H_{9/2}$ & 31.9 & 0.4459 & 0.08 & 0.4420 & 0.38 &
0.4385 \\ 
0.4277$^d$ & [Fe II] a$^4F_{7/2}$--a$^4G_{9/2}$ & 17.7 & & 2.00 & &
4.84 & \\  
0.4358$^d$ & [Fe II] a$^4F_{3/2}$--a$^4G_{5/2}$ & 11.2 && 1.29 & & 
3.12 & \\  
0.4359$^d$ & [Fe II] a$^6D_{7/2}$--a$^6S_{5/2}$ & 22.3 & & 2.36 
& & 6.04 & \\  
0.4356$^d$ & [Fe II] a$^4F_{5/2}$--b$^2P_{3/2}$ & 0.12 & & 
0.02 & & 0.08 & \\  
0.4411$^d$ & [Co II] a$^3F_2$--b$^3P_1$ & 5.93 & & 0.32 & & 0.18 & \\  
0.4414$^d$ & [Fe II] a$^6D_{5/2}$--a$^6S_{5/2}$ & 16.2 &  & 
1.72 & & 4.39 & \\  
0.4416$^d$ & [Fe II] a$^6D_{9/2}$--b$^4F_{9/2}$ & 23.6 & & 3.05 & & 7.92 & \\  
0.4658\footnote{Blend of two lines} & [Fe III] a$^5D_4$--$^3F2_4$ & 129 & 
0.4705 & 49.3 & 0.4715 & 87.7 & 0.4718 \\  
0.4702$^e$ & [Fe III]$^5D_3$--$^3F_3$ & 58.6 & & 25.7 & & 47.6 & \\  
0.4989\footnote {Blend of three lines} & [Fe III]$^5D_3$--$^3F_4$  
& --- & --- & 1.06 & 0.5032 & --- & --- \\  
0.5006$^f$ & [FeII] a$^4F_{5/2}$--b$^4F_{7/2}$ & --- & --- & 0.39 & & ---
& --- \\
0.5007$^f$ & [Fe II] a$^6D_{3/2}$--b$^4P_{5/2}$ & --- & --- & 0.20 & & --- 
& --- \\
0.5159\footnote {Blend of two lines} & [Fe II] a$^4F_{7/2}$--b$^4P_{3/2}$  
& 41.0 & 0.5305 & 6.96 & 0.5292 & 21.1 & 0.5295 \\  
0.5270$^g$ & [Fe III] a$^5D_3$--a$^3P2_2$ & 100 & & 36.0 & & 63.1 & \\  
0.5890\footnote {Blend of two lines} & [Co III]
a$^4F_{9/2}$--a$^2G_{9/2}$  
& 181 & 0.5930 & 34.6 & 0.5920 & 13.4 & 0.5920 \\  
0.5908$^h$ & [Co III]a$^4F_{7/2}$--a$^2G_{7/2}$ & 53.2 & & 11.8 & & 4.85
& \\  
0.6129\footnote {Blend of two lines} & [CoIII] a$^4F_{5/2}$--a$^2G_{7/2}$  
& 37.6 & 0.6218 & 8.36 & 0.6210 & 3.42 & 0.6195 \\  
0.6197$^i$ & [Co III] a$^4F_{7/2}$--a$^2G_{9/2}$ & 51.7 & & 9.88 & 
& 3.83 & \\  
0.6578 & [Co III] a$^4F_{9/2}$--a$^4P_{5/2}$ & 24.6 & 0.6622 
& 9.10 & 0.6620 & 5.28 & 0.6620 \\  
0.7155\footnote {Blend of two lines} & [Fe II]a$^4F_{9/2}$--a$^2G_{9/2}$  
& 31.2 & 0.7220 & 11.0 & 0.7195 & 52.8 & 
0.7205 \\  
0.7172$^j$ & [Fe II] a$^4F_{7/2}$--a$^2G_{7/2}$ & 10.9 & & 4.82 & & 
28.5 & \\  
0.7388\footnote {Blend of three lines} & [Fe II]
a$^4F_{5/2}$--a$^2G_{7/2}$ & 8.15 & 0.7565 & 3.59 & 0.7500 & 21.3 &
0.7470 \\  
0.7453$^k$ & [Fe II]a$^4F_{7/2}$--a$^2G_{9/2}$ & 9.59 & & 3.37 & & 16.2 &
\\  
0.7541$^k$ & [Co II]a$^3F_4$--a$^3P_2$ & 23.5 & & 2.96 & & 2.61 & \\  
0.8029 & [Co II]a$^3F_3$--a$^3P_1$ & 10.1 & 0.8150 & 1.29 & 0.8092 & ---
& --- \\
0.8617 & [Fe II] a$^4F_{9/2}$--a$^4P_{5/2}$ & --- & --- & 2.57 & 0.8662 &
16.3 & 
0.8680 \\  
\\ \hline 
\end{tabular} 
\end{minipage} 
\end{table*} 
\clearpage 
\begin{table*} 
\footnotesize 
\centering 
\begin{minipage}{160mm} 
\caption[]{Modelled Infrared Emission Features of SNe Spectra at t\(<\)100  
days} 
\centering 
\hspace{1cm} 
\begin{tabular}{clcccc} \hline \hline 
& &  
\multicolumn{2}{c} {SN1995D} &  
\multicolumn{2}{c} {SN1995al} \\  
& &  
\multicolumn{2}{c} {+92d} &  
\multicolumn{2}{c} {+94d} \\  
$\lambda_{rest}$\footnote {$\lambda$ is in microns} & 
\multicolumn{1}{c} {Identification} & 
Line Flux\footnote {Line flux is $\times$10$^{-15}$erg s$^{-1}$ cm$^{-2}$}   
& $\lambda_{peak}^a$ &  
Line Flux$^b$ & $\lambda_{peak}^a$ \\ \hline 
\\ 
0.9345\footnote{Blend of two lines} & [Co II] a$^3F_3$--a$^1D_2$ & 25.6 &
0.9570 & --- & --- \\
0.9531$^c$ & [S III] 3p$^2$ $^1D_2$--3p$^2$ $^3P_2$ & 80.0 & & --- & ---
\\  
1.0191\footnote{Blend of six lines} & [Co II] a$^3F_4$--b$^3F_4$ & 14.4 &
1.033 & 19.2 & 1.036  \\ 
1.0283$^d$ & [Co II] a$^3F_2$--b$^3F_2$ & 4.56 &  & 6.17 &  \\ 
1.0289$^d$ & [S II] 3p$^2$ $^2P_{3/2}$--3p$^3$ $^2D_{3/2}$ & 7.91 & &
13.9  & \\
1.0323$^d$ & [S II] 3p$^2$ $^2P_{3/2}$--3p$^3$ $^2D_{5/2}$ & 13.3 & &
23.4  & \\
1.0339$^d$ & [S II] 3p$^2$ $^2P_{1/2}$--3p$^3$ $^2D_{5/2}$  & 6.31 & &
10.8 & \\
1.0373$^d$ & [S II] 3p$^2$ $^2P_{1/2}$--3p$^3$ $^2D_{5/2}$ & 2.51 & &
4.30 & \\
1.2567\footnote {Blend of two lines} & [Fe II] a$^6D_{9/2}$--a$^4D_{7/2}$  
& 2.86 & 1.279  & 3.69 & 1.280 \\ 
1.2723$^e$ & [Co III] a$^4P_{5/2}$--a$^2D2_{5/2}$ & 1.39 & & 6.45  & \\ 
1.443 & [Fe I] a$^5D_4$--a$^5F_5$ & 1.71 & 1.452  & 3.30 & 1.449 \\ 
1.5474\footnote{Blend of two lines} & [Co II] a$^5F_5$--b$^3F_4$  
& 7.11 & 1.558  & 9.53 & 1.556 \\ 
1.5489$^f$ & [Co III] a$^2G_{9/2}$--a$^2H_{9/2}$ & 7.28  & & 29.1 &  \\ 
1.6435\footnote {Blend of two lines} & [Fe II] a$^4F_{9/2}$--a$^4D_{7/2}$  
& 2.11 & 1.649 & 2.72 & 1.646  \\ 
1.6348$^g$ & [Co II] a$^5F_1$--b$^3F_2$ & 1.57 & & 2.12 &  \\ 
1.7366\footnote{Blend of three lines} & [Co II] a$^5F_2$--b$^3F_3$ & 1.45
& 1.755  & 1.96 & 1.756  \\ 
1.7416$^h$ & [Co III] a$^2G_{9/2}$--a$^2H_{11/2}$ & 3.18 &  & 12.2 & \\ 
1.7643$^h$ & [Co III] a$^2G_{7/2}$--a$^2H_{9/2}$ & 1.92 & & 7.67 & \\ 
1.9571\footnote{Blend of three lines} & [Co III]
a$^4P_{1/2}$--a$^2P_{1/2}$  
& 2.95 & 2.008 & 11.8 & 1.997  \\ 
1.9810$^i$ & [Fe I] a$^5F_5$--a$^3F_4$ & 2.46 & & 3.26 & \\ 
2.0015$^i$ & [Co III] a$^4P_{5/2}$--a$^2P_{3/2}$ & 4.62 &  & 19.2 &  \\ 
2.0967\footnote{Blend of two lines} & [Co III] a$^4P_{3/2}$--a$^2P_{3/2}$  
& 2.36 & 2.110 & --- & --- \\ 
2.1457$^j$ & [Fe III] a$^3H_4$--a$^3G_3$ & 1.28 &  & --- & --- \\ 
2.2184\footnote{Blend of three lines} & [Fe III] a$^3H_6$--a$^3G_5$  &
2.36 &  2.248 & --- & --- \\
2.2427$^k$ & [Fe III] a$^3H_4$--a$^3G_4$ & 2.04 &  & --- & --- \\  
2.2799$^k$ & [Co III] a$^4P_{1/2}$--a$^2P_{3/2}$ & 0.89 & & --- & --- \\ 
\\ \hline 
\end{tabular} 
\end{minipage} 
\end{table*} 
\clearpage 
\begin{table*} 
\footnotesize 
\centering 
\begin{minipage}{160mm} 
\caption[]{Modelled Infrared Emission Features of SNe Spectra at t\(>\)100 
days} 
\centering 
\vspace{5mm} 
\begin{tabular}{clcccccc} \hline \hline 
& &  
\multicolumn{2}{c} {SN 1995al} & 
\multicolumn{2}{c} {SN 1994ae} &  
\multicolumn{2}{c} {SN 1991T} \\ 
& &  
\multicolumn{2}{c} {+176d} & 
\multicolumn{2}{c} {+175d} & 
\multicolumn{2}{c} {+363d} \\  
$\lambda_{rest}$\footnote{$\lambda$ is in microns} & 
\multicolumn{1}{c} {Identification} &  
Line Flux\footnote{Line flux is $\times$10$^{-15}$ erg s$^{-1}$ cm$^{-2}$}  
& $\lambda_{peak}^a$ &  
Line Flux$^b$ & $\lambda_{peak}^a$ & Line Flux$^b$ & $\lambda_{peak}^a$ \\ \hline 
\\ 
1.0191\footnote{Blend of six lines} & [Co II] a$^3F_4$--b$^3F_4$ & 2.61 &  
1.0355 & 2.48 & 1.030 & --- & ---   \\ 
1.0283$^c$ & [Co II] a$^3F_2$--b$^3F_2$ & 0.76  & & 0.71  & & --- & ---
\\ 
1.0289$^c$ & [S II] 3p$^2$ $^2P_{3/2}$--3p$^3$ $^2D_{3/2}$ & 3.19 & &
1.28  & & --- & --- \\
1.0323$^c$ & [S II] 3p$^2$ $^2P_{3/2}$--3p$^3$ $^2D_{5/2}$ & 5.34 & &
2.14 &  & --- & --- \\ 
1.0339$^c$ & [S II] 3p$^2$ $^2P_{1/2}$--3p$^3$ $^2D_{5/2}$ & 2.91 & &
0.51 & & --- & --- \\
1.0373$^c$ & [S II] 3p$^2$ $^2P_{1/2}$--3p$^3$ $^2D_{5/2}$ & 1.16  & &
0.58  & & --- & --- \\
1.2567\footnote{Blend of two lines} & [Fe II] a$^6D_{9/2}$--a$^4D_{7/2}$
& 1.84 & 1.269  & 2.21  & 1.264 & 14.6 & 1.267  \\ 
1.2723$^d$ & [Co III] a$^4P_{5/2}$--a$^2D2_{5/2}$ & 0.18 & & 0.04 &  &
0.05 &  \\
1.443 & [Fe I] a$^5D_4$--a$^5F_5$ & --- & ---  & 0.96 & 1.448 & 0.89 &
1.451  \\
1.5474\footnote{Blend of two lines} & [Co II] a$^5F_5$--b$^3F_4$ & 1.29  
& 1.555 & 1.23   & 1.550 & 1.52 & 1.547 \\ 
1.5489$^e$ & [Co III] a$^2G_{9/2}$--a$^2H_{9/2}$ & 1.53  & & 0.43 & &
0.63 & \\
1.6435\footnote{Blend of two lines} & [Fe II] a$^4F_{9/2}$--a$^4D_{7/2}$
&  1.36 & 1.651 & 1.63 & 1.649 & 10.8 & 1.654  \\ 
1.6348$^f$ & [Co II] a$^5F_1$--b$^3F_2$ & 0.26 & & 0.24 & & 0.29 & \\ 
1.7366\footnote{Blend of three lines} & [Co II] a$^5F_2$--b$^3F_3$ & 0.25  
& 1.754   & 0.23 & 1.749 & 0.28  & 1.812   \\ 
1.7416$^g$ & [Co III] a$^2G_{9/2}$--a$^2H_{11/2}$ & 0.87 &  & 0.31 &  &
0.47 & \\ 
1.7643$^g$ & [Co III] a$^2G_{7/2}$--a$^2H_{9/2}$ & 0.40 & & 0.11 & & 0.17
& \\ 
1.8000\footnote{Blend of two lines} & [Fe II] a$^4F_{5/2}$--a$^4D_{5/2}$  
& --- & --- & 0.31 & 1.8045 & 2.01 & \\ 
1.8094$^h$ & [Fe II] a$^4F_{7/2}$--a$^4D_{7/2}$ & --- & --- & 0.33 &  &
2.17 & \\ 
1.9571\footnote{Blend of three lines} & [Co III]
a$^4P_{1/2}$--a$^2P_{1/2}$  
& ---  & --- & 0.18 & 1.992 & --- & ---   \\ 
1.9810$^i$ & [Fe I] a$^5F_5$--a$^3F_4$ & --- & --- & 0.69 & & --- & --- \\ 
2.0015$^i$ & [Co III] a$^4P_{5/2}$--a$^2P_{3/2}$ & --- & ---  & 0.21 & &
--- & --- \\
2.1457 & [Fe III] a$^3H_4$--a$^3G_3$ & --- & --- & ---  & --- & 1.85 &
2.154 \\
2.2184\footnote{Blend of two lines} & [Fe III] a$^3H_6$--a$^3G_5$ & 1.38
& 3.243 & ---  & --- & 3.46 & 2.164 \\  
2.2427$^j$ & [Fe III] a$^3H_4$--a$^3G_4$ & 1.26  &  & ---  & --- & 3.09 &
\\  
2.3485 & [Fe III] a$^3H_5$--a$^3G_5$ & --- & --- & --- & --- & 2.15 &
2.362 \\ 
\\ \hline 
\normalsize 
\end{tabular} 
\end{minipage} 
\end{table*} 
\clearpage 
\newpage 
\begin{table*} 
\centering 
\begin{minipage}{160mm} 
\caption[]{Evolution of cobalt/iron mass ratio} 
\centering 
\vspace{5mm} 
\begin{tabular}{lcccccc} \hline \hline 
\multicolumn{1}{c}{Source} & Epoch & M$_{Co}$/M$_{Fe}$\footnote{Mass 
ratio values are as predicted by the $^{56}$Ni decay hypothesis. The 
errors indicate the range of mass ratio over which a satisfactory model fit 
to the data was achieved. Most of this was due to atomic data 
uncertainties.} \\  
& \multicolumn{1}{c}{(d)} \\ \hline 
\\ 
SN 1995D  & +92 & 0.66$\pm$0.36 \\ 
\\ 
SN 1995al & +94 & 0.66$^{+0.6}_{-0.3}$ \\ 
\\ 
SN 1995al & +176 & 0.25$^{+0.25}_{-0.125}$  \\ 
\\ 
SN 1994ae & +175 & 0.25$^{+0.7}_{-0.07}$  \\  
\\ 
SN 1991T & +338 & 0.04$\pm$0.015  \\ 
\\ 
\hline 
\end{tabular} 
\end{minipage} 
\end{table*} 
\begin{table*} 
\footnotesize 
\centering 
\begin{minipage}{160mm} 
\caption[]{Absolute mass estimates} 
\centering 
\vspace{5mm} 
\begin{tabular}{lcccccccccc} \hline \hline 
\multicolumn{1}{c}{Source} & Epoch & M$_{Fe^o}$ & M$_{Fe^+}$ &  
M$_{Fe^{2+}}$ & M$_{Fe^{3+}}$\footnote{The mass of  
triply-ionized material is determined using Axelrod's estimates.} & 
M$_{Co^o}$ & M$_{Co^+}$ & M$_{Co^{2+}}$ & M$_{Co^{3+}}$$^a$ & M$_{^{56}Ni}$ 
\footnote{The values for the total $^{56}$Ni mass are conservative  
as they are obtained by summing the singly- and doubly- ionized  
iron and cobalt only. Inclusion of neutral and triply-ionized species  
would increase the total mass by a factor of 1.5--2.5.}   \\ 
& \multicolumn{1}{c}{(d)}  
& \multicolumn{1}{c}{(M$_\odot$)} 
& \multicolumn{1}{c}{(M$_\odot$)} 
& \multicolumn{1}{c}{(M$_\odot$)} 
& \multicolumn{1}{c}{(M$_\odot$)} 
& \multicolumn{1}{c}{(M$_\odot$)} 
& \multicolumn{1}{c}{(M$_\odot$)} 
& \multicolumn{1}{c}{(M$_\odot$)} 
& \multicolumn{1}{c}{(M$_\odot$)} 
& \multicolumn{1}{c}{(M$_\odot$)} \\ \hline 
\\ 
SN 1995D & +92 & \(<\)0.040 & 0.040 & 0.120 & (0.022) & \(<\)0.027 & 
0.027 & 0.081 & (0.015) & 0.27$\pm$0.16 \\  
\\ 
SN 1995al & +94 & \(<\)0.040 & 0.040 & 0.32 & (0.044) & \(<\)0.027 & 0.027 &  
0.21 & (0.030) & 0.67$^{+1.1}_{-0.4}$ \\ 
\\ 
SN 1995al & +176 & --- & 0.025 & 0.22 & (0.078) & --- & 0.006 & 0.054 & 
(0.019)  
& 0.31$\pm$0.19 \\ 
\\ 
SN 1994ae & +174 & \(<\)0.013 & 0.026 & 0.090 & (0.040) & \(<\)0.003 &
0.006  & 0.022 & (0.010) & 0.14\footnote{Owing to the
low signal-to-noise the error here is a factor of 1.5 times larger than
for the other supernovae.}$^{+0.5}_{-0.05}$ \\
\\ 
SN 1991T & +363 & \(<\)0.0055 & 0.082 & 0.46 & (0.31) & \(<\)0.00026 &
0.004 & 0.022 & (0.015) & 0.57$\pm$0.14 \\ 
\\ 
\hline 
\normalsize 
\end{tabular} 
\end{minipage} 
\end{table*} 
\clearpage 
\bf{Figure Captions} \\ 
\normalsize 
 
Figure 1: Photospheric phase infrared spectra of SNe 1986G, 1992G, 1991T and 
1996X arranged by epoch, relative to t$_{Bmax}$ (see Table~2 for 
details). Each spectrum has been wavelength-shifted to the rest frame 
of the parent galaxy.  No reddening correction has been applied. For 
clarity, all of the spectra have been displaced vertically. The dotted 
horizontal lines on the left side indicate zero flux for each of the 
spectra. For SN~1991T zero flux is at the x-axis. The fluxes have also 
been multiplied by different factors (shown in brackets) to allow 
comparison of the features at the different epochs. 
 
Figure 2: Nebular phase infrared spectra of SNe 1991T, 1994ae, 1995D and 
1995al arranged by epoch.  Otherwise as described in Figure~1 caption. 
 
Figure 3: Optical spectra of SNe 1992G, 1991T, 1994ae, 1995D and 1995al 
arranged by epoch, relative to t$_{Bmax}$ (see Table~3 for details). 
For SN~1991T at +403~d zero flux is at the x-axis.  Other details as 
described in Figure~1 caption. 
 
Figure 4: Optical spectrum (solid line) of SN~1995D at +96 days, obtained 
at the Shane 3~m telescope of Lick Observatory (see Table~3 for details). 
It has not been wavelength-corrected for redshift, nor de-reddened. Also 
shown is our best model match (dashed line) (see Table~5) which has 
been red-shifted and reddened to match the SN. The same model parameters 
are used for the day~+92 NIR model spectrum shown in Fig.~5. Features we 
believe to be predominantly due to cobalt or iron are indicated. 
 
Figure 5: Near-infrared spectrum (solid line) of SN~1995D at +92 days, 
obtained at the United Kingdom Infrared Telescope with the cooled grating 
spectrometer CGS4 (see Table~2 for details). It has not been 
wavelength-corrected for redshift, nor de-reddened. Also shown is our 
best model match (dashed line) (see Table~5) which has been 
red-shifted and reddened to match the SN. The same model parameters are 
used for the optical model spectrum shown in Fig.~4. Features we believe 
to be predominantly due to cobalt, iron or sulphur are indicated. 
 
Figure 6: As Figure~5 but showing in detail the features in the 
wavelength range 1.5--2.5~$\mu$m. 
 
Figure 7: Near-infrared spectrum (solid line) of SN~1995al at +94 days, 
obtained at the United Kingdom Infrared Telescope with the cooled 
grating spectrometer CGS4 (see Table 2 for details). Other details as 
described in Figure 5 caption. 
 
Figure 8: Optical spectrum (solid line) of SN~1995al at +149 days, 
obtained with the intermediate dispersion spectrograph of the Isaac 
Newton Telescope, La Palma (see Table~3 for details). It has been scaled 
to correspond approximately to the epoch of the day~+176 NIR spectrum. 
The same model parameters are used for the day +176 NIR model spectrum 
shown in Fig.9. Other details as described in Figure 4 caption. 
 
Figure 9: Near-infrared spectrum (solid line) of SN~1995al at +176 days, 
obtained at the United Kingdom Infrared Telescope with the cooled 
grating spectrometer CGS4 (see Table~2 for details). Other details as 
described in Figure~5 caption. 

Figure 10: Near-infrared spectrum (solid line) of SN~1994ae at +175 days,
obtained at the United Kingdom Infrared Telescope with the cooled grating
spectrometer CGS4 (see Table~2 for details). To improve signal-to-noise,
the spectrum has been binned by a factor of $\times$4. Other details as
described in Figure~5 caption.

Figure 11: Optical spectrum (solid line) of SN~1991T at +403 days, 
obtained with the Cassegrain spectrograph on the Anglo-Australian 
Telescope (see Table~3 for details). The flux has been scaled to 
approximately the epoch of the +338~day near-infrared spectrum 
(Fig.~12). The same model parameters are used for the day~+338 NIR 
model spectrum shown in Fig.~12. Other details as described in Figure~4 
caption. 
 
Figure 12: Near-infrared spectrum (solid line) of SN~1991T at +338 days, 
obtained at the United Kingdom Infrared Telescope with the cooled 
grating spectrometer CGS4 (see Table~2 for details). Other details as 
described in Figure~5 caption. 
 
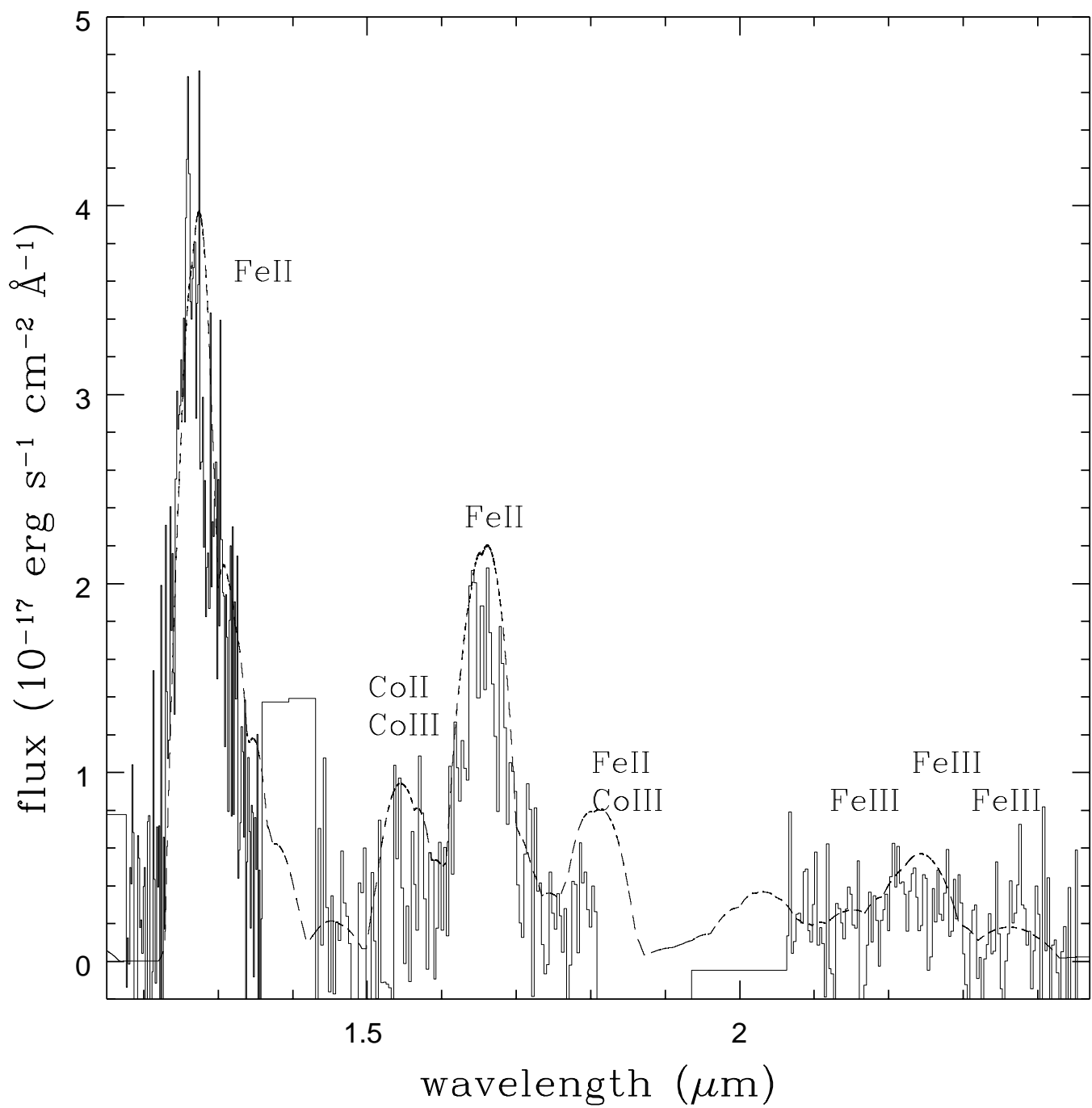
\end{document}